\documentclass[sigconf,screen]{acmart}
\AtBeginDocument{%
  }

\setcopyright{acmlicensed}
\copyrightyear{2026}
\acmYear{2026}
\setcopyright{cc}
\setcctype{by}
\acmConference[CHI '26]{Proceedings of the 2026 CHI Conference on Human Factors in Computing Systems}{April 13--17, 2026}{Barcelona, Spain}
\acmBooktitle{Proceedings of the 2026 CHI Conference on Human Factors in Computing Systems (CHI '26), April 13--17, 2026, Barcelona, Spain}
\acmPrice{}
\acmDOI{10.1145/3772318.3791848}
\acmISBN{979-8-4007-2278-3/2026/04}

\usepackage{multirow}
\usepackage{makecell}
\begin{document}

\title{Mind the Gap: Mapping Wearer–Bystander Privacy Tensions and Context-Adaptive Pathways for Camera Glasses}

%
\author{Xueyang Wang}
\orcid{0000-0002-9797-9491}
\email{wang-xy22@mails.tsinghua.edu.cn}
\affiliation{
    \institution{Tsinghua University}
    \city{Beijing}
    \country{China}
}

\author{Kewen Peng}
\orcid{0009-0000-3835-0127}
\email{astrid.peng@utah.edu}
\affiliation{
    \institution{University of Utah}
    \city{Salt Lake City, Utah}
    \country{United States}
}

\author{Xin Yi}
\authornote{Corresponding author.}
\orcid{0000-0001-8041-7962}
\email{yixin@tsinghua.edu.cn}
\affiliation{
    \institution{Tsinghua University}
    \city{Beijing}
    \country{China}
}
\affiliation{
    \institution{Beijing Academy of Artificial Intelligence}
    \city{Beijing}
    \country{China}
}

\author{Hewu Li}
\orcid{0000-0002-6331-6542}
\email{lihewu@cernet.edu.cn}
\affiliation{
    \institution{Tsinghua University}
    \city{Beijing}
    \country{China}
}

\renewcommand{\shortauthors}{Wang et al.}

\begin{abstract}
Camera glasses create fundamental privacy tensions between wearers seeking recording functionality and bystanders concerned about unauthorized surveillance. We present a systematic multi-stakeholder evaluation of privacy mechanisms through surveys (N=525) and paired interviews (N=20) in China. Study 1 quantifies expectation-willingness gaps: bystanders consistently demand stronger information transparency and protective measures than wearers will provide, with disparities intensifying in sensitive contexts where 65--90\% of bystanders would take defensive action. Study 2 evaluates twelve privacy-enhancing technologies, revealing four fundamental trade-offs that undermine current approaches: visibility versus disruption, empowerment versus burden, protection versus agency, and accountability versus exposure. These gaps reflect structural incompatibilities rather than inadequate goodwill, with context emerging as the primary determinant of privacy acceptability. We propose context-adaptive pathways that dynamically adjust protection strategies: minimal-friction visibility in public spaces, structured negotiation in semi-public environments, and automatic protection in sensitive contexts. Our findings contribute a diagnostic framework for evaluating privacy mechanisms and implications for context-aware design in ubiquitous sensing.
\end{abstract}




\maketitle

\section{Introduction}

\begin{figure*}[t]
  \centering
  \includegraphics[width=\linewidth]{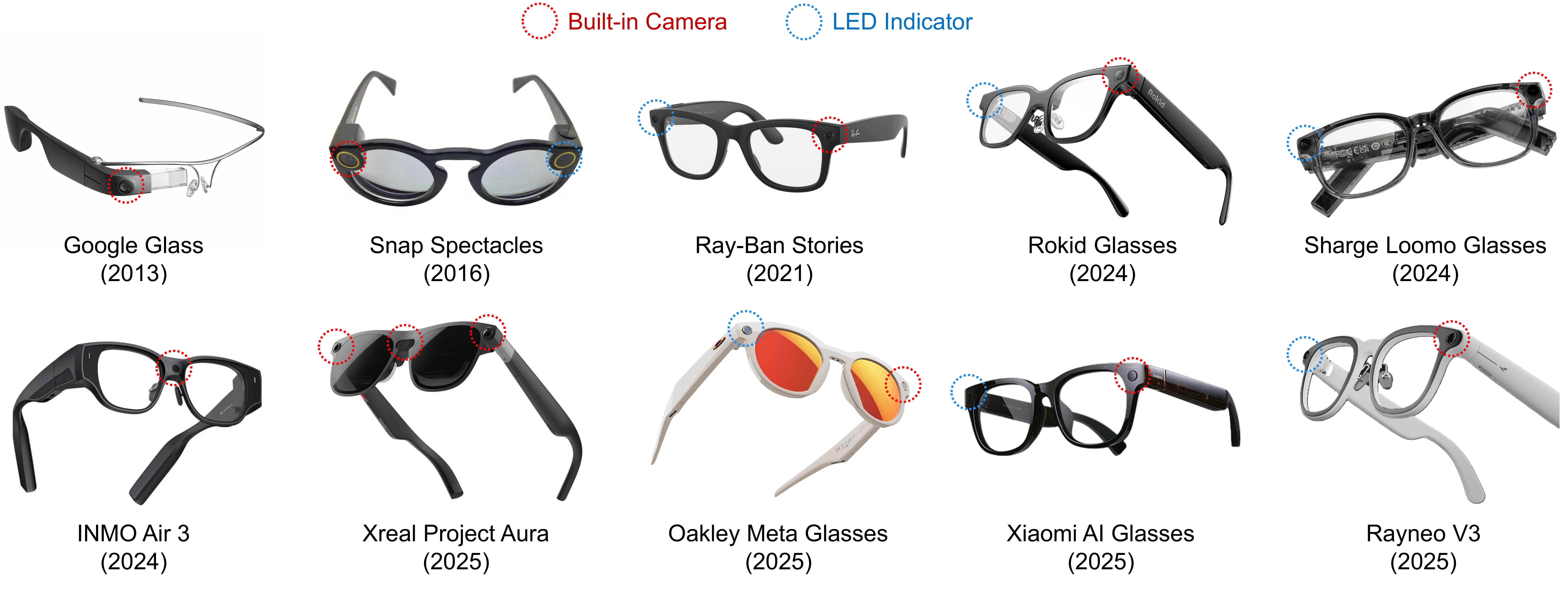}
  \caption{Representative camera-equipped smart glasses launched by major technology companies over the past decade. Their built-in cameras are often seamlessly disguised within everyday eyewear form factors, and LED indicators signaling recording status are typically small, making recording less perceptible to bystanders.}
  \label{fig:smartglasses-timeline}
\end{figure*}

Wearable camera glasses have rapidly transitioned from experimental prototypes to mainstream consumer products. Ray-Ban Meta Smart Glasses have surpassed 2 million units in global sales~\cite{vmagazineRayBanMeta}, while technology companies including Xiaomi, Rayneo, and Rokid have launched competing products (Figure~\ref{fig:smartglasses-timeline}). These devices offer compelling functionalities including voice assistants, real-time translation, and AR navigation, while embedding high-resolution cameras within fashionable eyewear designs~\cite{bipat2019analyzing}.

This integration creates fundamental privacy tensions absent in traditional photography. Unlike smartphones that require conspicuous gestures, camera glasses enable covert recording through voice commands or subtle touches~\cite{brookingsSeeingPast}. Current notification mechanisms prove inadequate: Ray-Ban Stories features only a small white LED that becomes imperceptible in bright environments and meaningless to unfamiliar observers~\cite{techcrunchFacebookWarned, bhardwaj2024focus, portnoff2015somebody}. Consequently, bystanders face unprecedented surveillance risks, including unauthorized facial recognition and AI-based inference~\cite{niu2025everyone, lebeck2018towards, o2023augmenting}.

Researchers have proposed various Privacy-Enhancing Technologies (PETs) to address these challenges~\cite{o2023privacy, de2019security, perez2017bystanders}. Wearer-side interventions include automatic bystander detection and blurring~\cite{david2024understanding, corbett2023bystandar} and enhanced recording indicators~\cite{bukhari2025rethinking, koelle2018beyond}. Bystander-controlled mechanisms enable recording refusal through wearable markers or gestures~\cite{koelle2018your, shu2017your, perez2018facepet}, but impose significant usability burdens in dynamic settings~\cite{wu2024designing}. Prior work has documented that wearers and bystanders hold different privacy expectations~\cite{denning2014situ, bhardwaj2024focus, o2023privacy}, and that multi-stakeholder conflicts arise across sensing technologies from smart homes to AR devices~\cite{ahmad2020tangible, thakkar2022would, chung2023negotiating}.

However, a critical gap persists: while existing work establishes that stakeholder conflicts exist, we lack systematic measurement of \textit{how much} expectations diverge across specific privacy dimensions, \textit{which} mechanisms might bridge these differences versus which face irreconcilable conflicts, and \textit{how} context moderates these gaps. Without such quantification, designers cannot prioritize interventions or anticipate where technical solutions will fail.

\begin{figure*}[t]
    \centering
    \includegraphics[width=0.9\linewidth]{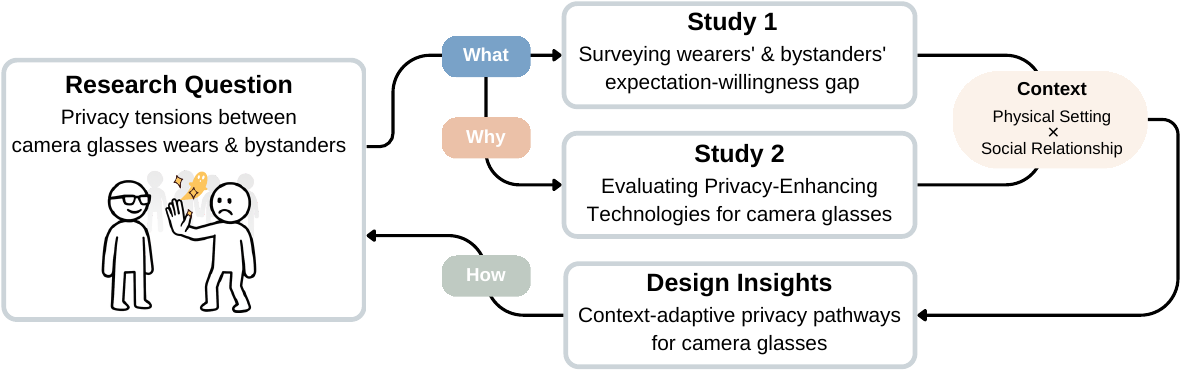}
    \caption{Research framework overview. Study 1 investigates the expectation-willingness gap between wearers and bystanders across contexts (\textit{what} is the gap). Study 2 evaluates existing PETs to understand underlying tensions (\textit{why} the gap persists). Design insights propose context-adaptive pathways (\textit{how} to address the gap).}
    \label{fig:overview}
\end{figure*}

To address this gap, we conducted a two-study investigation examining privacy perceptions from both stakeholder perspectives (Figure~\ref{fig:overview}). We situate our study in China, a rapidly growing smart glasses market ~\cite{xromForecasts107, xromXiaomiRaises}. While privacy norms vary across cultures, our findings reveal fundamental tensions likely relevant to other contexts ~\cite{rajaram2025privacy, abraham2024don, bhardwaj2024focus}. Study 1 employs a large-scale survey (N=525) to quantify privacy expectations across six contextual scenarios varying by physical setting and social relationship. We measured wearers' willingness to provide information transparency and protective measures against bystanders' expectations for these same dimensions. The results reveal persistent expectation-willingness gaps: bystanders demand significantly stronger data sharing control ($p<.01$) and prior consent ($p<.01$) than wearers will provide, with disparities intensifying in sensitive contexts where 65--90\% of bystanders would take defensive action.

Study 2 evaluates twelve representative PETs through paired interviews (N=20) combining HCI researchers' theoretical expertise with experienced users' practical insights. Systematic rating across effectiveness, usability, transparency, and social acceptability revealed four fundamental trade-offs: awareness mechanisms that inform bystanders inevitably disrupt social interactions (\textit{visibility versus disruption}), consent mechanisms empower bystanders by burdening them with self-defense (\textit{empowerment versus burden}), automated protection reduces user autonomy (\textit{protection versus agency}), and accountability requires privacy surrender through authentication (\textit{accountability versus exposure}).

These findings point toward a context-adaptive framework that operates through distinct pathways calibrated to environmental characteristics: minimal-friction visibility in public spaces, structured negotiation in semi-public environments, and automatic protection in sensitive contexts where vulnerability justifies reduced autonomy.

Our work makes three contributions to HCI research on ubiquitous sensing privacy:

\begin{itemize}
    \item \textbf{Quantification of expectation-willingness gaps.} Through parallel surveys of 525 wearers and bystanders, we provide systematic measurement of privacy expectation disparities across five information dimensions and five protective measures, revealing that gaps concentrate in control mechanisms and intensify in sensitive contexts.
    
    \item \textbf{Identification of fundamental trade-offs in PET design.} Through paired evaluation of 12 mechanisms, we identify four trade-offs (visibility versus disruption, empowerment versus burden, protection versus agency, accountability versus exposure) that explain why current approaches fail to reconcile stakeholder conflicts.
    
    \item \textbf{Context-adaptive privacy pathways.} Based on systematic preference patterns, we propose design pathways that select and combine mechanisms based on environmental characteristics, recognizing context rather than individual negotiation as the primary determinant of privacy acceptability.
\end{itemize}

\section{Related Works}

\subsection{Privacy Challenges Unique to Camera Glasses}

Wearable camera glasses have enabled seamless media capture through integrated cameras and on-device AI, supporting diverse applications from medical assistance~\cite{zhang2025can, chang2020medglasses} to navigation~\cite{nicholas2022friendscope} and learning~\cite{huh2025vid2coach}. However, this seamless experience creates distinctive privacy challenges between wearers and bystanders~\cite{iqbal2023adopting, koelle2015don, tran2025wearable}.

Unlike web cookies where consent is discrete and individual, camera glasses privacy is situated and dynamic, emerging through ongoing information exchange. Compared with traditional cameras or head-mounted displays (HMDs) like VR/AR headsets, camera glasses appear ordinary yet enable continuous recording, intensifying tensions between everyday visibility and pervasive capture. Traditional photography relies on visible gestures (e.g., raising a phone) that signal recording intent~\cite{nguyen2009encountering, procyk2014exploring}. In contrast, camera glasses enable capture through subtle gestures or voice commands~\cite{tran2025wearable, chung2023negotiating}, leading bystanders to assume continuous recording~\cite{koelle2015don}. Although many devices include LED indicators, such cues often remain imperceptible at distance, easily obstructed, or socially ambiguous~\cite{tran2025wearable, portnoff2015somebody, koelle2018beyond}.

Beyond recording ambiguity, smart glasses transform privacy risks into an instantaneous process through AI-based recognition~\cite{iqbal2023adopting}. Real-time facial recognition and scene understanding provide users with assistance while simultaneously turning bystanders into live data sources before they can become aware or exercise control. This shifts privacy threats from post-hoc content review to real-time inference, where sensitive attributes such as demographics, health status, or affiliations could be extracted~\cite{niu2025everyone, lebeck2018towards, gopal2023hidden}. These technical characteristics converge to undermine established social norms for negotiating photography~\cite{kudina2019ethics}. Traditional photo-taking was visually marked and normatively negotiable, enabling bystanders to object or opt out. Smart glasses render such negotiation difficult: wearers may use them as ordinary eyewear without realizing potential offense, while bystanders lack cues to interpret or contest recording~\cite{bhardwaj2024focus, denning2014situ}. This becomes particularly problematic in sensitive environments such as fitting rooms and medical facilities~\cite{rashidi2018you, nguyen2011situating}. These unresolved tensions have repeatedly triggered public backlash and market withdrawals~\cite{corbett2023securing, krombholz2015ok}.

We summarize these characteristics in Table~\ref{tab:comparison}. Collectively, these challenges illustrate that privacy in camera glasses requires re-establishing contextual and social mechanisms for negotiation~\cite{liao2019understanding, vimalkumar2021okay}, motivating our examination of context-aware and multi-stakeholder approaches.

\begin{table*}[t]
\centering
\caption{Comparison of privacy-relevant characteristics across recording devices. HMDs refer to head-mounted displays with visible form factors. Camera glasses refer to AI-enabled eyewear resembling ordinary glasses.}
\label{tab:comparison}
\small
\setlength{\tabcolsep}{5pt}
\renewcommand{\arraystretch}{1.3}
\begin{tabular}{@{}p{3.2cm}p{3.2cm}p{2.3cm}p{2.8cm}p{2.3cm}@{}}
\toprule
\textbf{Dimension} & \textbf{Privacy Relevance} & \textbf{Camera/Phone} & \textbf{HMD} & \textbf{Camera Glasses} \\
\midrule
\textbf{Recording ambiguity} 
& How easily bystanders detect capture intent~\cite{nguyen2009encountering} 
& High (gestures) 
& Medium (appearance) 
& Low (subtle) \\
\addlinespace[0.2em]
\textbf{Real-time AI processing} 
& Extent of automated recognition at capture~\cite{iqbal2023adopting, rashidi2018you, nguyen2011situating} 
& Low (manual) 
& Medium (task-specific) 
& High (continuous) \\
\addlinespace[0.2em]
\textbf{Established social norms} 
& Clarity of shared expectations for recording~\cite{bhardwaj2024focus, denning2014situ} 
& High (clear consent)
& Medium (emerging) 
& Low (ambiguous) \\
\bottomrule
\end{tabular}
\end{table*}

\subsection{Contextual Factors in Privacy Expectations}
\label{2.2}

Privacy has long been understood as a dynamic boundary-regulation process rather than a fixed state~\cite{windl2022automating}. Altman's theory framed privacy as ongoing negotiation between desired and actual levels of access to the self~\cite{altman1975environment}, and subsequent research has consistently shown that privacy expectations shift with where data is captured and who is involved~\cite{altman1977privacy, joinson2010privacy, egelman2015predicting}. This pattern is particularly salient in emerging technologies like camera glasses, where users report heightened concern about information being repurposed beyond immediate contexts~\cite{gallardo2023speculative}.

Nissenbaum's theory of Contextual Integrity (CI) offers a normative lens for understanding these dynamics~\cite{nissenbaum2004privacy}. According to CI, privacy violations arise when information flows breach contextual norms of appropriateness (what may be revealed) or distribution (how it may circulate). These norms are shaped by three components: \textit{contextual conditions} that define a setting, the \textit{social roles of actors}, and the \textit{transmission principles} that regulate information flows. For camera glasses, transmission principles for ubiquitous capture remain undefined, making their establishment both urgent and central to our work.

\textbf{Physical setting} has emerged as among the most influential dimensions for understanding camera glasses privacy. Settings structure the visibility and permeability of information flows, influencing whether recording behaviors are perceived as contextually appropriate~\cite{altman1977privacy, nissenbaum2004privacy}. Prior research documents that privacy expectations differ significantly across spatial contexts, with tolerance typically higher in public spaces and lower in private or sensitive ones~\cite{denning2014situ, o2023privacy}. These expectations reflect situated norms: shared understandings about appropriate behavior within a given space.

Equally critical is the \textbf{social relationship} between actors~\cite{windl2025designing, corbett2023securing, lebeck2018towards}. Prior work in adjacent domains shows that relationship strength (e.g., friend, colleague, stranger) systematically shapes disclosure comfort and privacy expectations~\cite{wiese2011you, stutzman2010friends}. Relationship determines the degree of trust and legitimacy perceived in data capture~\cite{olson2005study, fogel2009internet, akter2020uncomfortable}. With the growing popularity of camera glasses, redefining boundaries between wearers and bystanders has become urgent.

Despite the centrality of context to privacy, prior work has largely examined contextual factors in isolation~\cite{denning2014situ, akter2020uncomfortable}. O'Hagan et al.~\cite{o2023privacy} systematically varied context in evaluating bystander attitudes toward AR sensing, finding strong effects of both setting and relationship. Windl et al.~\cite{windl2023understanding} examined technology-facilitated privacy violations across physical contexts. However, these studies primarily capture \textit{single-stakeholder} perspectives. Our work extends this line by systematically comparing how the same contextual factors differentially shape expectations for wearers versus bystanders, revealing where gaps emerge and intensify.

\subsection{Stakeholder Perspectives and Privacy Negotiation}
\label{2.3}

Camera glasses create asymmetric privacy relationships where wearers control recording while bystanders bear exposure risks~\cite{chung2023negotiating}. This asymmetry transforms privacy into continuous negotiation between primary users and secondary actors who often interpret the same recording behavior differently~\cite{koelle2015don}.

\textbf{Bystander-focused research} has extensively documented concerns about consent and surveillance. Denning et al.~\cite{denning2014situ} conducted in-situ studies revealing bystanders' discomfort with AR glasses recording. Subsequent work has explored bystander awareness needs~\cite{ahmad2020tangible, pierce2022addressing}, defensive responses~\cite{zhao2023if, singhal2016you}, and protection mechanisms~\cite{hasan2020automatically, jimenez2014tag, al2025bystandaria}. These studies establish that bystanders consistently desire stronger notification and control than current devices provide.

\textbf{Wearer-focused research} has examined different concerns. Bhardwaj et al.~\cite{bhardwaj2024focus} interviewed camera glasses wearers about their privacy dilemmas, finding that many wearers do consider bystander perspectives but face practical constraints in addressing them. Bipat et al.~\cite{bipat2019analyzing} analyzed camera glasses use in the wild, documenting usage patterns and social challenges. Tran et al.~\cite{tran2025wearable} surveyed wearers about notification preferences, finding general willingness to signal recording but concerns about social friction.

However, this separation of stakeholder perspectives leaves critical gaps. First, it remains unclear \textit{how much} bystander expectations diverge from wearer willingness across specific privacy dimensions. Second, without direct comparison, we cannot identify which mechanisms might bridge these differences versus which face irreconcilable conflicts. Third, the interaction between stakeholder role and context remains underexplored. 

\textbf{Multi-stakeholder approaches} have begun addressing these limitations. Chung et al.~\cite{chung2023negotiating} examined dyadic interactions through AR glasses, revealing negotiation dynamics but focusing on acquainted pairs. Windl et al.~\cite{windl2025designing} designed consent mechanisms for spontaneous AR interactions, incorporating both user and target perspectives. Abraham et al.~\cite{abraham2024don} explored how sensitive contexts shape both wearer and bystander attitudes toward AR sensing. In smart home contexts, parallel work has examined owner-bystander tensions around domestic cameras~\cite{thakkar2022would, pierce2022addressing, wu2024designing}, demonstrating that multi-stakeholder conflicts are pervasive across ubiquitous sensing technologies.

Our work builds on and extends this foundation in several ways. First, we \textbf{quantify} the expectation-willingness gap through large-scale comparative measurement (N=525), enabling precise identification of where stakeholder requirements diverge most sharply. Prior work has documented that gaps exist; we measure their magnitude across specific dimensions and contexts. Second, we systematically evaluate whether existing Privacy-Enhancing Technologies (PETs) can \textbf{bridge} these measured gaps, revealing fundamental trade-offs that explain why current approaches fail. Third, we derive empirically-grounded implications for \textbf{context-adaptive pathways} that dynamically adjust protection strategies based on the contextual patterns our data reveal. This progression from gap quantification through mechanism evaluation to design recommendations represents a more complete treatment than prior single-study approaches.

\subsection{Privacy-Enhancing Technologies for Bystanders}\label{2.4}

\subsubsection{Sensor Transparency and Recording Notification Mechanisms}

Recording notifications seek to establish informed consent by enabling bystanders to detect when smart glasses capture audio or video. Current implementations rely primarily on visual indicators—Snap Spectacles employ circular LED rings \cite{koelle2018beyond}, while Ray-Ban Meta features a ``Capture LED'' during recording. However, these minimal indicators suffer from fundamental limitations. LEDs assume constant bystander vigilance yet fail when individuals are distracted, facing away, or have sensory impairments \cite{portnoff2015somebody, bhardwaj2024focus}. Research confirms that visual indicators alone prove inadequate for conveying device activity awareness \cite{ahmad2020tangible}.

Multimodal notification designs address these limitations by combining visual, auditory, and digital channels. Recent proposals include smartphone notifications to nearby devices and context-sensitive audio alerts \cite{bukhari2025rethinking, pierce2022addressing, aditya2016pic}. Empirical studies demonstrate user preference for multimodal approaches in privacy-sensitive contexts, though implementation faces inherent tensions between noticeability and obtrusiveness \cite{thakkar2022would, koelle2018beyond}. Smart home research provides relevant precedents through "tangible privacy" mechanisms \cite{ahmad2020tangible} and privacy visualization systems \cite{prange2021priview, albayaydh2023examining}, yet camera glasses' mobility demands solutions optimized for spontaneous encounters rather than controlled environments.

\subsubsection{Empowering Bystander Control and Refusal}

Beyond awareness, bystanders seek active intervention capabilities to signal recording refusal \cite{perez2020user, marky2020you}. One approach employs personal countermeasures: FacePET uses infrared LEDs to blind cameras \cite{perez2018facepet}, while InPhysible camouflages physiological signals \cite{mcduff2018inphysible}. Despite experimental effectiveness, these solutions prove impractical by requiring bystanders to carry specialized devices in anticipation of encounters.

Environmental interventions offer broader coverage through infrastructure-based solutions. BlindSpot proposed jamming signals in sensitive spaces \cite{patel2009blindspot}, while systems like I-Pic and Cardea enable wireless privacy preference broadcasting \cite{aditya2016pic, shu2016cardea}. These approaches eliminate individual equipment burdens but require industry-wide protocol adoption and regulatory enforcement currently absent from the ecosystem.

Marker-based consent signaling represents a third approach, enabling individuals to wear visual identifiers or perform gestures that trigger automatic recording cessation \cite{bo2014privacy, shu2017your, koelle2018your}. However, practical deployment faces critical obstacles: bystanders must anticipate risks and prepare markers, visible indicators may compromise user privacy, and enforcement depends on universal manufacturer compliance. Without mandatory standards, non-compliant devices can simply ignore signals, limiting these solutions to conceptual or prototype stages.

\subsubsection{Automated Bystander Privacy Protection}
Automated approaches implement privacy protection directly within recording devices through computer vision and signal processing. BystandAR distinguishes interaction targets from bystanders using eye tracking and spatial audio \cite{corbett2023bystandar}, while other systems employ real-time face blurring \cite{alharbi2019mask, hasan2020automatically} or activity recognition in degraded images \cite{dimiccoli2018mitigating}. Contextual factors including location \cite{roesner2014world}, social connections \cite{williamson2022digital}, and accessibility needs \cite{wolf2018we} can drive automatic control decisions. PrivacEye exemplifies sophisticated approaches by using eye movement analysis to detect privacy sensitivity and trigger mechanical shutters \cite{steil2019privaceye}.

Despite technical advances, automated protection faces fundamental limitations. Bystanders remain unaware of post-processing applications, perpetuating trust deficits regardless of actual protection levels. Recent research highlights risks of overlooking legitimate privacy concerns due to inadequate bystander definitions \cite{niu2025not}. Effective deployment requires coordinated ecosystem development including transparency mechanisms, industry standards, and public education, suggesting that technical solutions alone cannot resolve the complex sociotechnical challenges of ubiquitous sensing.

\section{Study 1: Surveying Wearers' and Bystanders' Privacy Expectations}
\label{study1}

To systematically understand privacy expectations surrounding camera glasses, we conducted a large-scale survey examining perspectives from two stakeholder groups: current or potential smart glasses users (\textit{wearers}, N = 232) and individuals potentially affected by such devices (\textit{bystanders}, N = 293). Our scenario-based design evaluated privacy attitudes across situations varying by \textbf{physical setting} (public, semi-public, private/sensitive spaces) and \textbf{social relationship} (strangers versus acquaintances). This dual-perspective approach enables direct comparison of expectations and identification of conflicts between groups. Study 1 addresses two research questions:

\begin{itemize}
\item \textbf{RQ1:} How do contextual factors (physical setting and social relationship) influence wearers' and bystanders' privacy concerns and behavioral intentions?
\item \textbf{RQ2:} What gaps exist between bystanders' expectations and wearers' willingness regarding information transparency and protective measures? How effectively do current notification mechanisms meet stakeholder needs?
\end{itemize}

\subsection{Survey Design}

\subsubsection{Contextual Scenario Design}
Grounded in Contextual Integrity theory~\cite{nissenbaum2004privacy}, we designed six vignettes systematically varying across physical setting and social relationship in a 3$\times$2 design, balancing comprehensiveness with participant fatigue~\cite{o2023privacy, windl2023understanding}:

\begin{itemize}
\item \textbf{Public spaces:} Street (travel vlog with companions), Mall (shopping among strangers)
\item \textbf{Semi-public spaces:} Meeting room (documentation with colleagues), Hospital (consultation among strangers)
\item \textbf{Private/sensitive spaces:} Private party (casual recording with friends), Gym (workout documentation among strangers)
\end{itemize}

Each vignette portrayed typical camera glasses use from the participant's assigned role (wearer or bystander), with presentation order randomized using Latin square counterbalancing to reduce sequencing bias~\cite{lazar2017research, eisenberg1988order}.

\subsubsection{Technology Primer}
To address varying familiarity levels across participants, all respondents received a standardized technology primer before evaluation. This primer included: (1) representative product images and brand examples (Ray-Ban Meta, Xiaomi AI Glasses), (2) typical use cases and core functions, (3) annotated photographs showing LED indicator placement, and (4) explanation of LED signaling conventions. This ensured all participants possessed sufficient baseline knowledge for informed responses regardless of prior experience.

\subsection{Measures}

\subsubsection{Demographics and Baseline Attitudes}
We collected demographic information including gender, age, education, camera glasses familiarity, and brand awareness. To establish baseline privacy attitudes, we employed validated scales adapted for each stakeholder group (see Appendix ~\ref{appendix:survey} for complete items).

For \textbf{bystanders}, we adapted the Internet Users' Information Privacy Concerns (IUIPC) scale~\cite{malhotra2004internet}, measuring three dimensions with two items each: \textit{Awareness} (need for disclosure about data collection), \textit{Control} (perceived control over information practices), and \textit{Collection} (concerns about unauthorized recording).

For \textbf{wearers}, we developed scales capturing other-regarding privacy attitudes. Drawing on the Privacy Orientation Scale~\cite{baruh2014more}, we created two items measuring \textit{Perceived Responsibility} toward bystanders. We then used Protection Motivation Theory (PMT)~\cite{rogers1975protection, maddux1983protection} as a generative lens to design items capturing \textit{Privacy Protection Intention} (willingness to modify behavior when others object) and \textit{Information Sharing Intention} (willingness to inform bystanders). We chose PMT over value-oriented scales such as VOPP~\cite{hasan2023psychometric} because our goal was to assess context-specific behavioral intentions rather than general privacy values. While PMT was originally developed for self-protective behaviors, its threat/coping appraisal structure extends naturally to other-regarding contexts, an approach established in healthcare informatics research on protecting patient privacy~\cite{ma2015survey, lee2021structural}.

Given overall survey complexity (six scenarios $\times$ multiple dimensions), we made a deliberate measurement trade-off: baseline attitudes used abbreviated 2-item scales for descriptive purposes, while scenario-based measures, which support our core claims, received greater emphasis.

\subsubsection{Contextual Measures}
For each scenario, we collected role-specific measures. Bystanders evaluated: privacy concerns (PC(B)), behavioral responses (BH(B)), information needs (I(B)), and protective measure expectations (PT(B)). Wearers assessed: recording reasonability (R(W)), concerns about affecting bystanders (PC(W)), information disclosure willingness (I(W)), and protective measure willingness (PT(W)).

Following prior literature on privacy negotiation~\cite{denning2014situ, chung2023negotiating, o2023privacy, hoyle2014privacy, gallardo2023speculative}, we operationalized five information dimensions and five protective mechanisms:

\begin{itemize}
\item \textbf{Information dimensions:} Purpose (intended use), Content (recording details), Sharing (upload/distribution), Retention (storage duration), AI Use (recognition/analysis)
\item \textbf{Protective measures:} Proactive Notification, Privacy Filter (face blurring), No Sharing (upload prohibition), Auto Delete, Prior Consent
\end{itemize}

Both used 7-point matrix-style Likert scales, supplemented by open-ended questions.

\subsubsection{LED Indicator Evaluation}
Participants evaluated existing LED indicators on: adequacy (5-point scale), inadequacy reasons (multiple choice), preferred notification methods (multiple choice), and adoption motivators (multiple choice).

\subsubsection{Scale Development and Validation}
Our instrument development followed a rigorous multi-stage process~\cite{lazar2017research}. First, two researchers independently collected candidate items from relevant literature~\cite{malhotra2004internet, rogers1975protection, denning2014situ, o2023privacy}, then collaboratively merged similar items and removed semantically ambiguous or construct-inconsistent items. Second, two external experts in cybersecurity and privacy reviewed the instrument to refine wording, ensure non-leading phrasing, and verify accessibility for general audiences. Third, we conducted three iterative pilot rounds (N=5 each), asking participants to identify ambiguous statements and checking for ceiling/floor effects. Feedback was incorporated until no ambiguities were reported.

\subsection{Participants and Procedure}

We recruited participants from mainland China through distinct channels. Bystanders were recruited via university mailing lists using materials broadly describing the study as examining attitudes toward wearable electronics to avoid response bias. Wearers were recruited through smart glasses enthusiast groups, brand communities, and forums. China provided an ideal research context as the world's largest smart glasses market---IDC projects 2.75 million units shipped in 2025 (107\% year-over-year growth)~\cite{xromForecasts107}, with brands like Xiaomi achieving rapid adoption~\cite{xromXiaomiRaises}.

Participants completed the 8-minute survey for USD \$1 compensation after providing informed consent. Quality controls excluded responses with completion times under 3 minutes, Mahalanobis $D^{2}$ outlier detection for straightlining, and failed attention checks, yielding 293 valid bystander responses (18.6\% exclusion rate) and 232 valid wearer responses (13.4\% exclusion rate).

Sample characteristics reflect typical technology adoption patterns (see Table ~\ref{tab:demographics} in Appendix). Wearers exhibited early adopter profiles with higher male representation (59.5\% vs. 49.2\%), greater device experience (23.3\% vs. 4.4\% current/former users), and increased awareness of international brands (Ray-Ban Meta: 43.5\% vs. 27.0\%). Both groups showed high familiarity with domestic brands, particularly Xiaomi AI Glasses (>87\%). Including participants with varying familiarity levels reflects real-world market conditions where most potential bystanders are non-users; our goal was to evaluate privacy mechanisms rather than device usability. The substantial sample sizes provide robust coverage across the adoption spectrum, though high education levels (>80\% bachelor's degree) may limit broader generalizability.

The study protocol was reviewed and approved by the Institutional Review Board (IRB) of Tsinghua University, and we strictly protected participants’ data privacy throughout the study.

\subsection{Measurement Validity}

Our scenario-based scales demonstrated strong internal consistency: Information Needs (bystanders, $\alpha$ = 0.89), Protective Measure Expectations (bystanders, $\alpha$ = 0.89), Information Disclosure Willingness (wearers, $\alpha$ = 0.94), and Protective Measure Willingness (wearers, $\alpha$ = 0.93). Baseline attitude scales showed modest reliability typical of 2-item measures ($\alpha$ = 0.48--0.77), appropriate for their descriptive purpose.

To assess convergent validity, we examined correlations between baseline measures and scenario-averaged outcomes (see Figure ~\ref{fig:validity_correlations} in Appendix). For bystanders, baseline items correlated positively with scenario-based privacy concerns and protection demands (r $\approx$ .12--.42, most ps < .001). For wearers, baseline responsibility and intention items correlated moderately with scenario-based disclosure willingness and protection adoption (r $\approx$ .25--.62, ps < .001). These patterns support treating abbreviated baseline scales as valid descriptive measures, though we acknowledge their limitations for primary analyses. Therefore, we treat these short baseline scales as secondary descriptive measures and do not rely on them as primary evidence for our core claims.

Exploratory factor analyses confirmed construct validity for all scenario-based measures. Each construct yielded a dominant first factor explaining 38.9--61.9\% of variance, with item loadings ranging from 0.56 to 0.86 and item-total correlations from 0.50 to 0.83, supporting their use as composite measures.

\begin{figure*}[t]
\centering
\includegraphics[width=\textwidth]{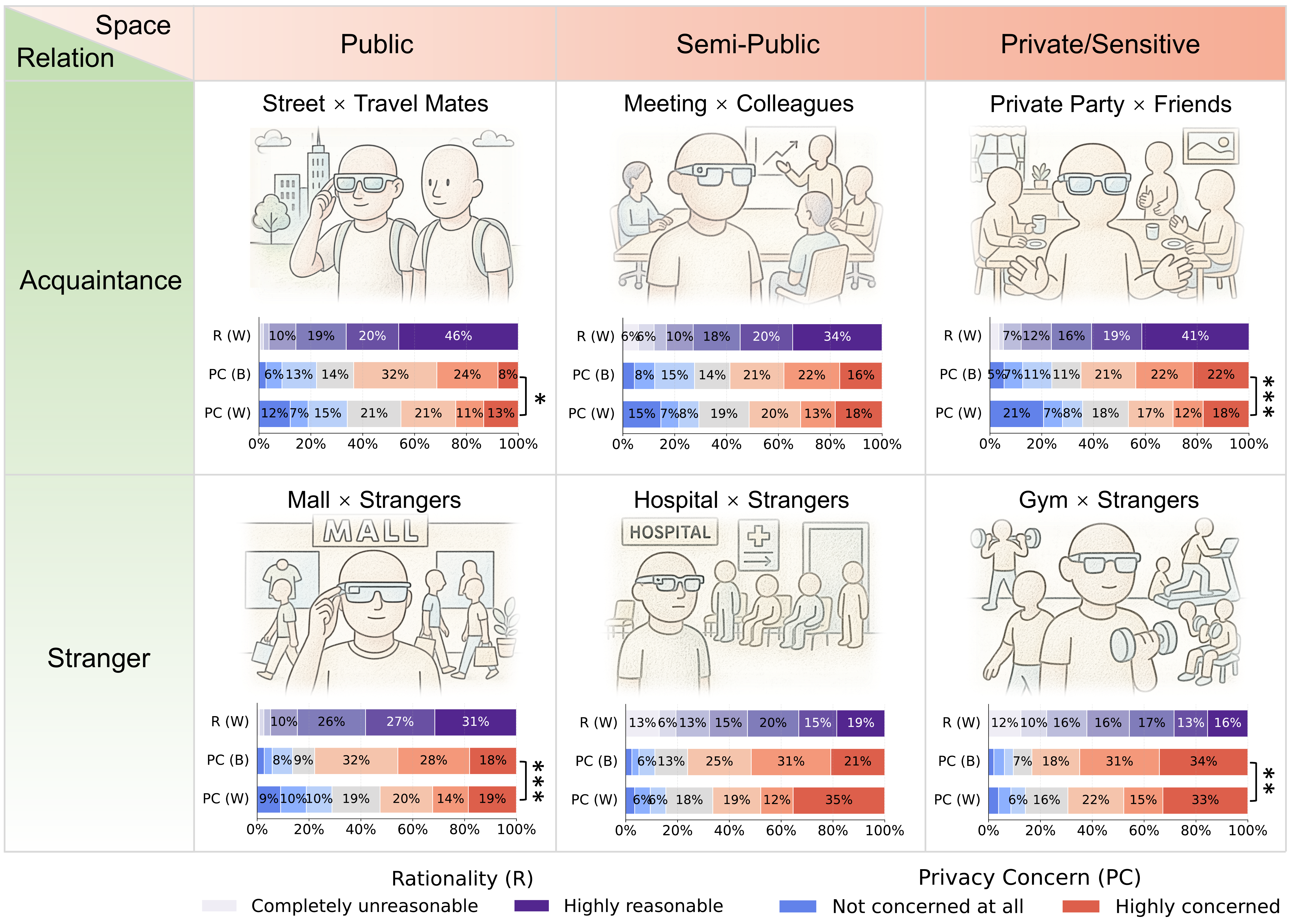}
\caption{Privacy perceptions across six contextual scenarios. Each panel shows scenario illustration and distribution of responses for Recording Reasonability (R(W)) by wearers, Privacy Concern by bystanders (PC(B)), and Privacy Concern by wearers (PC(W)). Asterisks indicate significant group differences (* $p<.05$, ** $p<.01$, *** $p<.001$).}
\label{fig:study-1-scenes}
\end{figure*}

\subsection{Statistical Analysis}

We employed the Aligned Rank Transform (ART) procedure~\cite{wobbrock2011aligned} to accommodate non-normal Likert-scale distributions. Mixed-design ANOVAs examined Group (between-subjects: wearer vs. bystander) and Scenario (within-subjects: 6 vignettes) effects on privacy concerns, information disclosure needs/willingness, and protective measure expectations/willingness. Post-hoc comparisons used ART-C~\cite{elkin2021aligned} with Holm corrections.

\subsection{Results}

\subsubsection{Recording Reasonability and Privacy Concerns}

Wearers consistently viewed recording as reasonable across all contexts, with mean ratings exceeding the neutral midpoint even in sensitive settings (Street $\times$ Travel Mates: $M=5.92$, $SD=1.26$; Gym $\times$ Strangers: $M=4.17$, $SD=1.94$). A significant Scenario effect ($F=59.548$, $p<.001$) revealed clear hierarchies: public recordings with companions were deemed most reasonable, while medical and fitness contexts were least reasonable though still above neutral (Figure~\ref{fig:study-1-scenes}).

Bystanders expressed significantly higher privacy concerns than wearers across all scenarios ($F=20.540$, $p<.001$; bystanders: $M=5.07$, $SD=1.59$; wearers: $M=4.59$, $SD=1.93$). Privacy-sensitive spaces---Gym $\times$ Strangers ($M=5.46$) and Hospital $\times$ Strangers ($M=5.26$)---triggered the highest concerns ($F=56.440$, $p<.001$). Group $\times$ Scenario interactions revealed persistent perception gaps, with bystanders reporting significantly higher concerns in four scenarios: Street $\times$ Travel Mates ($\Delta M=0.51$, $p<.05$), Private Party $\times$ Friends ($\Delta M=0.79$, $p<.001$), Mall $\times$ Strangers ($\Delta M=0.72$, $p<.001$), and Gym $\times$ Strangers ($\Delta M=0.42$, $p<.01$). These disparities suggest wearers systematically underestimate privacy implications, particularly in social and commercial settings.

\subsubsection{Behavioral Response Patterns}

\begin{figure*}[t]
\centering
\includegraphics[width=\textwidth]{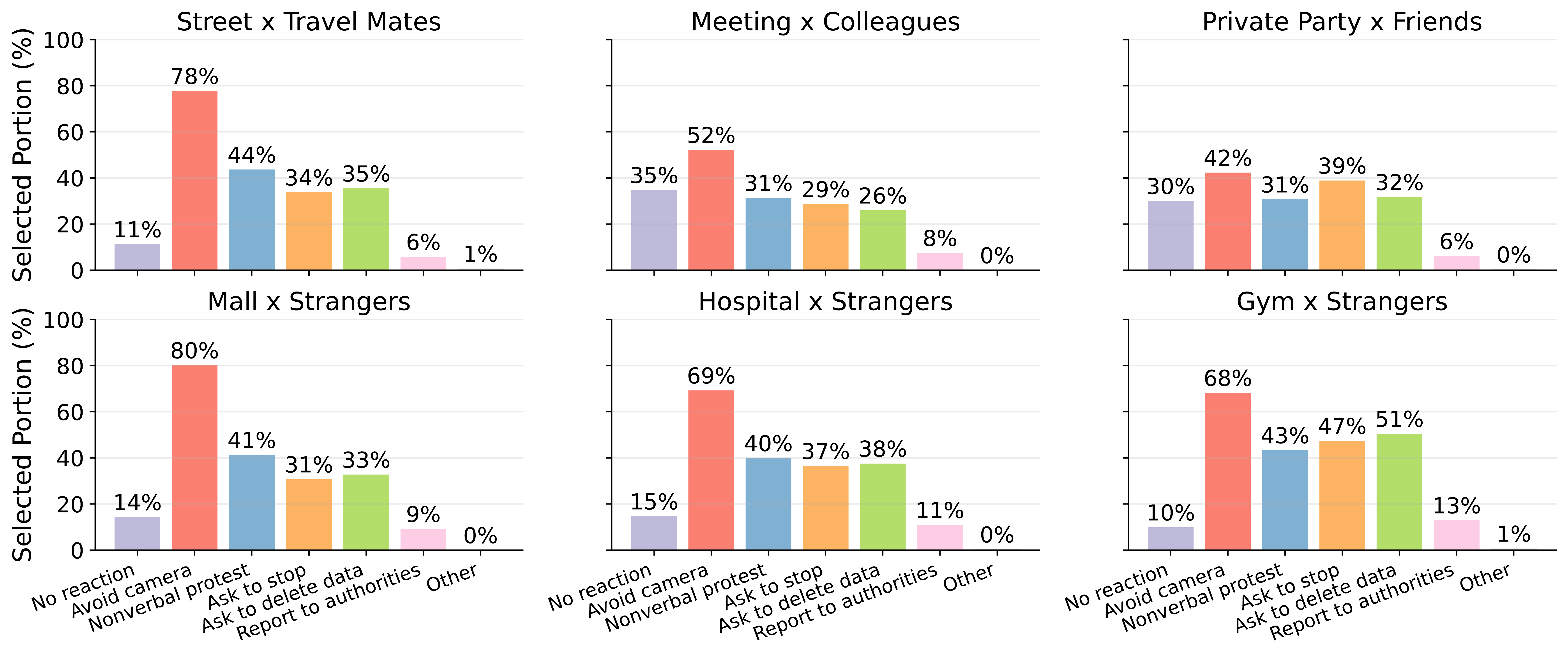}
\caption{Bystanders' anticipated behavioral responses to smart glasses recording across six scenarios. Bars represent the percentage of participants endorsing each response option (multiple selections allowed).}
\label{fig:study-1-behaviors}
\vspace{5mm}
\end{figure*}

Bystanders demonstrated strong defensive intentions when encountering smart glasses recording (Figure~\ref{fig:study-1-behaviors}). Camera avoidance dominated across contexts (52--80\%), peaking in commercial spaces (Mall: 80\%) and fitness facilities (Gym: 68\%) where anonymity expectations are highest. Responses followed a clear escalation hierarchy: nonverbal protests (31--51\%) served as middle ground, while direct interventions, requesting recording cessation (26--51\%) or data deletion (31--39\%), increased in unfamiliar settings. Formal complaints to authorities, though rare (0--13\%), peaked in medical (11\%) and fitness (13\%) environments.

\begin{figure*}[t]
\centering
\includegraphics[width=\textwidth]{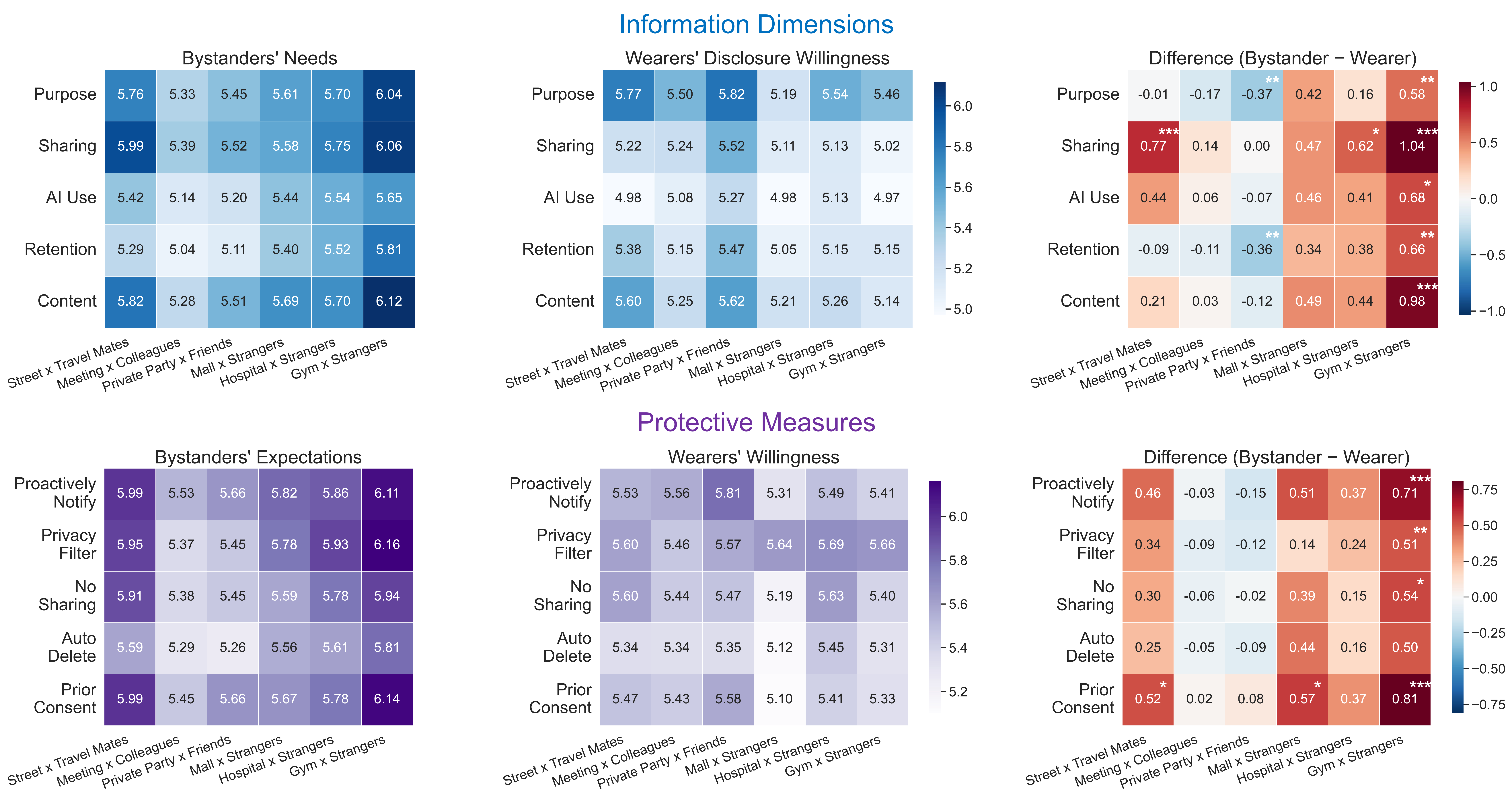}
\caption{Information transparency and protective measure expectations across scenarios. \textbf{Top row:} Information dimensions. \textbf{Bottom row:} Protective measures. Left panels show bystanders' needs, middle panels show wearers' willingness, and right panels show the difference (Bystander $-$ Wearer). Warmer colors indicate larger gaps. Asterisks denote significant differences (* $p<.05$, ** $p<.01$, *** $p<.001$).}
\label{fig:study-1-info-protect}
\end{figure*}

\begin{figure*}[t]
\centering
\includegraphics[width=\textwidth]{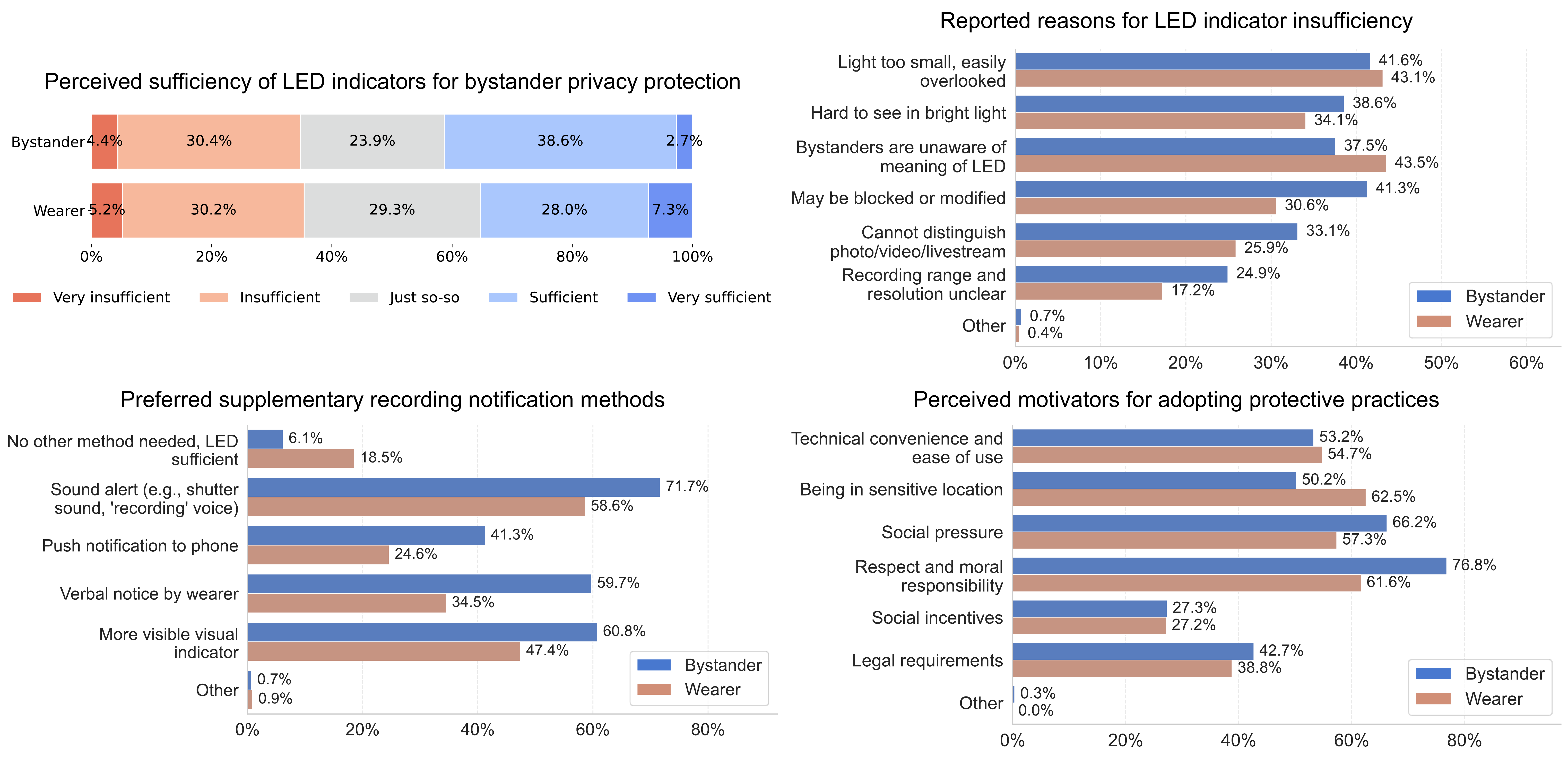}
\caption{Evaluation of LED indicators and preferences for enhanced recording notifications. \textbf{Top left:} Perceived sufficiency of LED indicators. \textbf{Top right:} Reasons for insufficiency. \textbf{Bottom left:} Preferred additional notification methods. \textbf{Bottom right:} Perceived motivators for adopting privacy-protective practices. Blue bars represent bystanders; red bars represent wearers.}
\label{fig:study-1-LEDs}
\end{figure*}

Passive acceptance remained uncommon (10--15\%) except in professional settings (35\%) where workplace dynamics may discourage resistance. With 65--90\% of bystanders indicating they would take defensive action, these findings underscore the need for proactive privacy mechanisms that prevent rather than trigger defensive behaviors.

\subsubsection{Information Transparency and Protection Expectations}

Systematic gaps emerged between bystanders' expectations and wearers' willingness across information transparency and protective measures (Figure~\ref{fig:study-1-info-protect}). Two dimensions showed significant group differences: bystanders demanded stricter \textbf{data sharing} control than wearers would provide ($F=10.233$, $p<.01$), and \textbf{prior consent} expectations significantly exceeded wearers' willingness ($F=7.835$, $p<.01$).

Context amplified these disparities systematically. Gym $\times$ Strangers produced the largest gaps across multiple dimensions: data sharing ($\Delta M=1.04$, $p<.001$), recording purpose ($\Delta M=0.58$, $p<.001$), content transparency ($\Delta M=0.98$, $p<.001$), prior consent ($\Delta M=0.81$, $p<.001$), and proactive notification ($\Delta M=0.71$, $p<.001$). Hospital $\times$ Strangers showed similar patterns, while familiar settings exhibited smaller disparities. This mirrors privacy concern findings---contexts triggering heightened concerns generate demands for transparency and protection that wearers are unwilling to meet.

\subsubsection{Adequacy of Current Recording Indicators}

Neither stakeholder group viewed LED indicators as adequate privacy protection (Figure~\ref{fig:study-1-LEDs}). Only 41.3\% of bystanders and 35.3\% of wearers considered LEDs sufficient. Participants identified critical failure modes: LEDs are too small and easily overlooked (41.6\% bystanders, 43.1\% wearers), become invisible in bright environments (38.6\%, 34.1\%), remain meaningless to unfamiliar observers (37.5\%, 43.5\%), and can be deliberately obstructed (41.3\%, 30.6\%).

Participants strongly endorsed multi-modal notification systems. Sound alerts garnered highest support (71.7\% bystanders, 58.6\% wearers), followed by enhanced visual indicators (60.8\%, 47.4\%). Role-dependent preferences emerged: bystanders favored systemic solutions like smartphone notifications (41.3\% vs.\ 24.6\%), while wearers preferred interpersonal approaches like verbal notice (59.7\% vs.\ 34.5\%). Only 6.1\% of bystanders and 18.5\% of wearers believed LEDs alone sufficed.

Beyond technical solutions, participants identified moral responsibility as the primary driver for privacy protection (76.8\% bystanders, 61.6\% wearers), followed by social pressure to avoid conflict (66.2\%, 57.3\%) and contextual sensitivity (50.2\%, 62.5\%). Legal requirements (42.7\%, 38.8\%) and social incentives (27.3\%, 27.2\%) played secondary roles, suggesting that fostering privacy-protective behaviors requires appealing to ethical sensibilities rather than regulatory compliance.

\subsubsection{Role of Familiarity with Smart Glasses}
\label{subsec:familiarity}

\begin{figure*}[t]
\centering
\includegraphics[width=\linewidth]{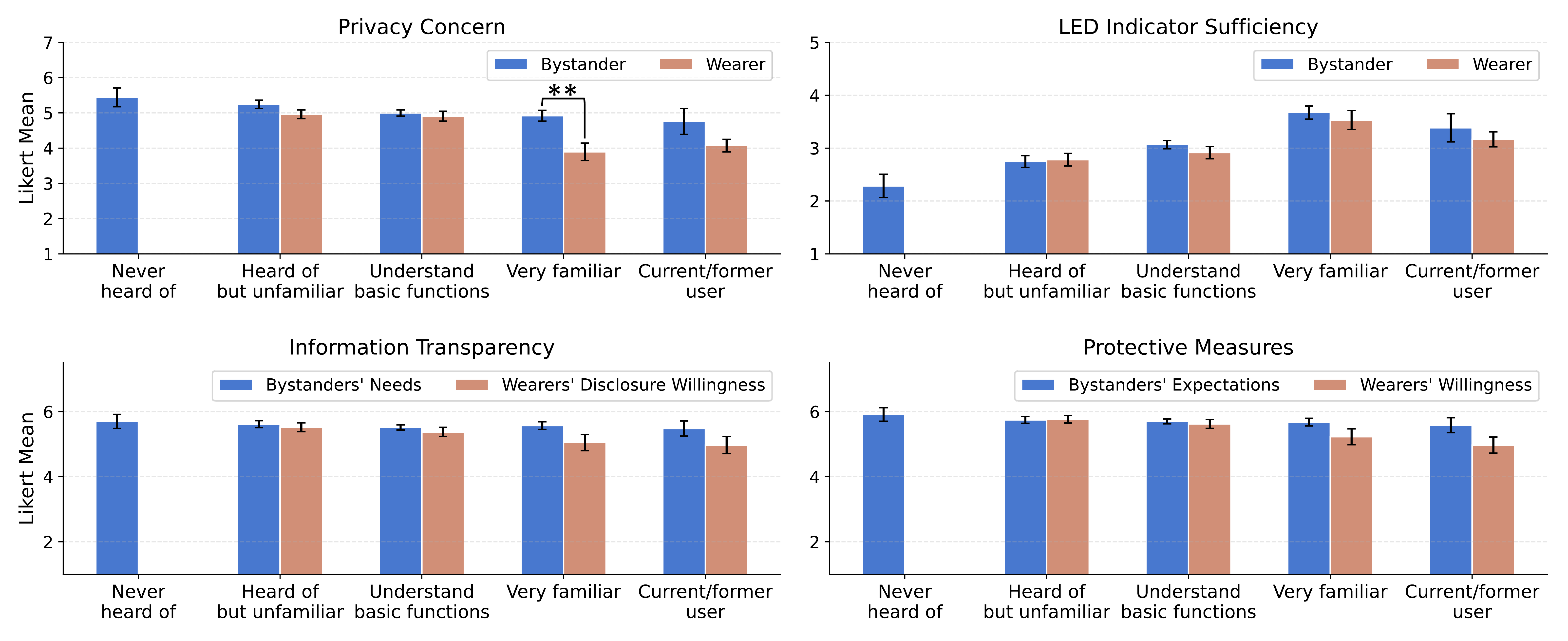}
\caption{Effects of familiarity with smart glasses on privacy attitudes. Each panel shows mean ratings (error bars: 95\% CI) for bystanders (blue) and wearers (orange) across four familiarity levels. Note: no participant in wearers group is ``never heard of camera glasses''.}
\label{fig:familiarity_four_metrics}
\end{figure*}

To examine whether our findings are driven by participants unfamiliar with smart glasses, we analyzed how self-reported familiarity relates to scenario-averaged privacy attitudes. We conducted ANOVAs with group (bystander vs.\ wearer) and familiarity as factors (Figure~\ref{fig:familiarity_four_metrics}).

For \textbf{privacy concern}, we observed significant main effects of group ($F(1,503)=14.87$, $p<.001$) and familiarity ($F(3,503)=9.94$, $p<.001$), and critically, a Group $\times$ Familiarity interaction ($F(3,503)=3.44$, $p=.017$). Bystanders maintained consistently high concerns across all familiarity levels, while wearers' concerns decreased with experience, widening the gap between groups. For \textbf{LED sufficiency}, perceived adequacy increased with familiarity ($F(3,503)=8.77$, $p<.001$) but remained around the scale midpoint even among current users, with no group differences. For \textbf{information transparency} and \textbf{protective measures}, neither showed significant effects of group, familiarity, or their interaction (all $p$s > .12).

These results demonstrate that including unfamiliar participants does not artificially inflate our findings. Bystanders' expectations remain uniformly high regardless of familiarity, while the primary effect is that experienced wearers become less concerned yet still do not view LEDs as sufficient. This reinforces our claim that privacy gaps are structural rather than artifacts of unfamiliarity, consistent with prior work showing that familiarity does not resolve privacy concerns in ubiquitous computing contexts~\cite{al2021role}.

\subsection{Summary of Study 1 Findings}

Our examination of privacy perceptions across six contextual scenarios reveals fundamental misalignments between stakeholder groups that illuminate why current camera glasses struggle with social acceptance.

\subsubsection{RQ1: Context as Primary Determinant of Privacy Acceptability}

Physical settings and social relationships emerge as primary determinants of privacy expectations, but with asymmetric effects across groups~\cite{o2023privacy, rajaram2025exploring}. Wearers maintained relatively stable perceptions of recording reasonability across all contexts (means > 4.0), operating under an assumption of general acceptability. Bystanders, however, exhibited strong contextual sensitivity: privacy-sensitive spaces triggered not only heightened concerns but also defensive behavioral intentions---up to 80\% would take action in commercial and fitness contexts.

This asymmetry reveals a fundamental disconnect: wearers view context as modulating the \textit{degree} of acceptable recording, while bystanders experience context as determining \textit{whether} recording should occur at all. The consistency of defensive responses (65--90\% across scenarios) indicates that current designs trigger conflict rather than facilitate negotiation. Privacy emerges as a situated phenomenon where spatial norms, social dynamics, and power relationships converge to define acceptable sensing practices~\cite{bhardwaj2024focus, o2023augmenting}.

\subsubsection{RQ2: The Systematic Expectation-Willingness Gap}

Our findings reveal a persistent chasm between bystanders' expectations and wearers' willingness, most acute for \textbf{data sharing control} and \textbf{prior consent}, precisely the mechanisms that would provide bystanders meaningful agency~\cite{windl2025designing}. The disparity intensifies in sensitive contexts where bystanders' vulnerability peaks while wearers' willingness plateaus, creating an inverse relationship that exacerbates tensions.

This gap reflects structural incompatibilities rather than inadequate goodwill. Bystanders' demands represent reasonable responses to involuntary surveillance~\cite{lebeck2018towards, niu2025everyone}, while wearers' reluctance reflects practical constraints of device functionality~\cite{wu2024designing}. Current notification mechanisms exemplify this incompatibility: LED indicators fail both groups---deemed inadequate by two-thirds of participants---due to limitations in perceptibility, interpretability, and circumvention resistance, consistent with prior work~\cite{portnoff2015somebody, koelle2018beyond}.

\subsubsection{Implications for Privacy Mechanism Design}

The expectation-willingness gap cannot be resolved through better individual mechanisms or user education alone~\cite{thakkar2022would}. Contextual variation demands adaptive rather than universal approaches, while inadequate notifications reveal fundamental tensions between transparency and covert recording capabilities~\cite{krombholz2015ok}. The prominence of moral responsibility as a motivator (outweighing technical convenience or legal requirements) suggests that effective solutions must engage ethical frameworks and social norms, pointing toward environmental protections that remove the burden from individual bystanders~\cite{shu2016cardea, abraham2024don}.

The persistent defensive responses indicate that passive awareness mechanisms often create rather than resolve privacy conflicts, calling for a shift from post-capture remediation to prevention-oriented design. Study 2 investigates whether existing PETs can bridge these structural gaps.

\section{Study 2: Evaluating Privacy-Enhancing Technologies for Camera Glasses}
\label{study2}

Study 1 revealed systematic expectation-willingness gaps between wearers and bystanders that current privacy mechanisms fail to address. Bystanders consistently demanded stronger transparency and control than wearers were willing to provide, with disparities intensifying in sensitive contexts. LED indicators failed both stakeholder groups, while defensive behavioral intentions (65--90\% of bystanders) indicated that current approaches generate rather than resolve privacy conflicts.

These findings raise critical questions about whether existing Privacy-Enhancing Technologies (PETs) can bridge these structural misalignments. To investigate this, Study 2 evaluates twelve representative PETs through paired interviews combining HCI researchers/designers (who contribute theoretical expertise) with experienced smart glasses users (who provide practical insights). This dyadic approach enables assessment of both technical feasibility and real-world viability. Study 2 addresses two research questions:

\begin{itemize}
    \item \textbf{RQ3: Context-Dependent Viability.} Can PETs adapt to the contextual variation in privacy needs identified in Study 1, or are new paradigms necessary?
    \item \textbf{RQ4: Effectiveness-Usability Trade-offs.} Do current PETs successfully balance privacy protection with practical usability?
\end{itemize}

\subsection{PET Selection and Classification}

We conducted a literature review following PRISMA guidelines~\cite{page2021prisma} to identify privacy protection mechanisms for bystanders across camera glasses, AR/VR devices, and smart home technologies. We queried SCOPUS, ACM Digital Library, and IEEE Xplore, targeting premier HCI venues (CHI, IMWUT, CSCW), AR/VR conferences (IEEE VR, ISMAR), and privacy forums (IEEE S\&P, SOUPS, PoPETs). Using keywords related to bystander privacy, wearable cameras, and privacy-enhancing technologies, we identified 127 unique records. Two researchers independently screened papers based on: (1) explicit bystander privacy focus, (2) proposed technical mechanisms, and (3) wearable camera relevance. This yielded 70 relevant papers, supplemented by commercial implementations from Snap Spectacles, Ray-Ban Meta, and Apple Vision Pro (Cohen's $\kappa$ = 0.83).

Through thematic synthesis~\cite{thomas2008methods} and reference to existing taxonomies~\cite{perez2017bystanders, de2019security}, we organized PETs into four functional categories:

\begin{itemize}
\item \textbf{Wearer-side Awareness Mechanisms (W):} Signal recording status through visual, auditory, or digital channels
\item \textbf{Bystander-side Consent Mechanisms (B):} Enable privacy preference expression through gestures, markers, broadcasting, or negotiation platforms
\item \textbf{Context-aware Automatic Processing (C):} Automatically protect privacy through face anonymization or location-based restrictions
\item \textbf{Platform-level Accountability Systems (P):} Ensure traceability through watermarking and post-hoc notifications
\end{itemize}

We selected 12 representative PETs spanning all categories with varying implementation complexity (Table~\ref{tab:PETs}). Selection prioritized: (1) comprehensive category coverage, (2) existence of functional prototypes or commercial deployments, and (3) potential to address the expectation-willingness gaps identified in Study 1.

\begin{table*}[t]
\centering
\caption{Privacy-Enhancing Technologies (PETs) Evaluated in Study 2}
\label{tab:PETs}
\begin{tabular}{p{2.2cm}lp{5.2cm}p{4.5cm}}
\toprule
\textbf{Category} & \textbf{PET Name} & \textbf{Description} & \textbf{Literature Examples} \\
\midrule

\multirow{12}{2.2cm}{\textbf{Wearer-side Awareness (W)}} 
& W1: LED Ring Indicator  
& Circular LED array around camera lens; different colors indicate recording modes (white: video, green: AI, orange: livestream).
& Snap Spectacles~\cite{spectaclesMessages}; Visual indicators~\cite{bhardwaj2024focus, koelle2018beyond, ahmad2020tangible, bukhari2025rethinking} \\
\cmidrule{2-4}

& W2: Audio Alerts 
& Speaker emits shutter sounds for photos and verbal announcements for videos.
& Korean camera standard~\cite{wiredKoreaBeeping}; Privacy Speaker~\cite{thakkar2022would} \\
\cmidrule{2-4}

& W3: External Display  
& Front-facing e-ink or LED display showing recording status to bystanders.
& Apple Vision Pro EyeSight~\cite{eyesightAVP}; MirrorCam~\cite{koelle2019evaluating}; EyeCam~\cite{teyssier2021eyecam} \\
\cmidrule{2-4}

& W4: Proximity Broadcast  
& Glasses broadcast recording status via BLE/WiFi to nearby smartphones.
& WiFi notifications~\cite{pidcock2011notisense}; BLE transparency~\cite{escher2022transparency}; PriView~\cite{prange2021priview} \\

\midrule

\multirow{12}{2.2cm}{\textbf{Bystander-side Consent (B)}} 
& B1: Gesture Recognition 
& Camera recognizes standardized gestures: open palm (stop) or thumbs-up (consent).
& Social signal detection~\cite{meirose2018towards, koelle2018your}; Gesture opt-out~\cite{shu2017your, alharbi2019mask} \\
\cmidrule{2-4}

& B2: Wearable Markers  
& Bystanders wear IR emitters, QR-coded clothing, or ultrasonic beacons signaling ``do not record.''
& FacePET~\cite{perez2018facepet}; Visual tags~\cite{bo2014privacy}; Beacons~\cite{ashok2014not, liao2025bystander} \\
\cmidrule{2-4}

& B3: Preference Broadcasting 
& Smartphone app broadcasts privacy preferences via BLE to nearby devices.
& I-Pic~\cite{aditya2016pic}; Cardea~\cite{shu2016cardea}; iRYP~\cite{sun2020iryp}; Do Not Capture~\cite{ra2017not} \\
\cmidrule{2-4}

& B4: Negotiation Platform 
& Real-time permission requests sent to bystander phones with allow/deny options.
& Erebus~\cite{kim2023erebus}; Interactive negotiation~\cite{zhou2024bring, li2016privacycamera} \\

\midrule

\multirow{6}{2.2cm}{\textbf{Context-aware Automatic Processing (C)}} 
& C1: Face Anonymization  
& AI-powered detection automatically blurs unauthorized faces during recording.
& Bystander detection~\cite{hasan2020automatically}; BystandAR~\cite{corbett2023bystandar}; PrivacEye~\cite{steil2019privaceye} \\
\cmidrule{2-4}

& C2: Geofencing Control 
& GPS/WiFi-based system automatically disables recording in sensitive zones.
& PlaceAvoider~\cite{templeman2014placeavoider}; World-driven access~\cite{roesner2014world} \\

\midrule

\multirow{4}{2.2cm}{\textbf{Platform Accountability (P)}} 
& P1: Digital Watermarking 
& Recording embeds immutable watermarks with device ID and timestamp.
& Geo-tagged media~\cite{henne2013snapme} \\
\cmidrule{2-4}

& P2: Face Matching 
& Platforms notify pre-registered users when their faces appear in uploaded content.
& HideMe~\cite{li2019hideme}; Cloak~\cite{zhang2018cloak, shu2016cardea} \\

\bottomrule
\end{tabular}
\vspace{5mm}
\end{table*}

\subsection{Participants and Procedure}

We recruited 20 participants forming 10 dyadic pairs (Table~\ref{tab:study2_participants}): 10 HCI researchers/designers with privacy and wearable technology expertise, and 10 camera glasses users with at least three months of device experience (recruited from Study 1 respondents). This pairing enables evaluation from both theoretical and practical perspectives. Sessions lasted approximately two hours with \$30 USD compensation.

\begin{table*}[t]
\centering
\caption{Study 2 Participant Demographics. Exp. = years of experience (HCI) or duration of device usage (Users).}
\label{tab:study2_participants}
\resizebox{0.8\textwidth}{!}{%
\begin{tabular}{c|ccccc|ccccc}
\toprule
\textbf{S.} & \multicolumn{5}{c|}{\textbf{HCI Researcher/Designer (H)}} & \multicolumn{5}{c}{\textbf{Camera Glasses User (U)}} \\
& PID & Age & Gen. & Specialization & Exp. & PID & Age & Gen. & Occupation & Exp. \\
\midrule
1 & PH1 & 25 & M & HCI Research & 2y & PU1 & 41 & M & Business Manager & 9mo \\
2 & PH2 & 27 & M & Usable Privacy & 3y & PU2 & 27 & M & Business Manager & 1y \\
3 & PH3 & 23 & F & HCI Design & 1y & PU3 & 29 & M & Business Manager & 1.5y \\
4 & PH4 & 23 & F & HCI Design & 5y & PU4 & 30 & F & Company Employee & 3mo \\
5 & PH5 & 23 & F & HCI Design & 4y & PU5 & 20 & M & Company Employee & 2mo \\
6 & PH6 & 24 & F & Usable Privacy & 3y & PU6 & 29 & F & Business Manager & 7mo \\
7 & PH7 & 22 & F & HCI Research & 2y & PU7 & 23 & M & Company Employee & 6mo \\
8 & PH8 & 23 & F & HCI Research & 2y & PU8 & 33 & M & Civil Servant & 3mo \\
9 & PH9 & 21 & M & Usable Privacy & 3y & PU9 & 30 & M & Business Manager & 6.5y \\
10 & PH10 & 28 & F & Platform Accountability & 4y & PU10 & 28 & M & Graduate Student & 3mo \\
\bottomrule
\end{tabular}
}
\end{table*}

Each session followed a four-phase protocol. In \textbf{Phase 1} ($\sim$10 min), facilitators introduced study objectives and established rapport through warm-up questions. In \textbf{Phase 2} ($\sim$60 min), participants examined 12 PETs supported by textual descriptions, literature images, scenario diagrams, and UI prototypes (see supplementary materials). Each participant independently rated mechanisms on four 7-point dimensions: \textit{Privacy Protection Effectiveness}, \textit{User Experience and Convenience}, \textit{Transparency and Trust}, and \textit{Social Acceptability and Scalability}. Dyads then discussed evaluations and proposed improvements; presentation order was randomized. In \textbf{Phase 3} ($\sim$40 min), participants re-evaluated PETs across three environmental categories (public, semi-public, private/sensitive spaces) using scenario cards, ranking mechanisms by contextual suitability. \textbf{Phase 4} ($\sim$10 min) captured additional insights through synthesis and reflection.

The study protocol was reviewed and approved by the Institutional Review Board (IRB) of Tsinghua University, and we strictly protected participants’ data privacy throughout the study.

\subsection{Analysis}

Interview recordings were transcribed verbatim and analyzed using grounded theory principles~\cite{mccann2003grounded}. Two researchers independently conducted open coding~\cite{blandford2016qualitative} to identify emergent themes around PET effectiveness, usability concerns, contextual appropriateness, and implementation barriers. Through iterative review, we developed a codebook (see supplementary materials) and re-analyzed transcripts using axial coding~\cite{mayring2014qualitative} to examine patterns across dyads. Inter-rater reliability achieved Cohen's $\kappa$ = 0.79, with disagreements resolved through discussion.

\begin{figure*}[t]
\centering
\includegraphics[width=\textwidth]{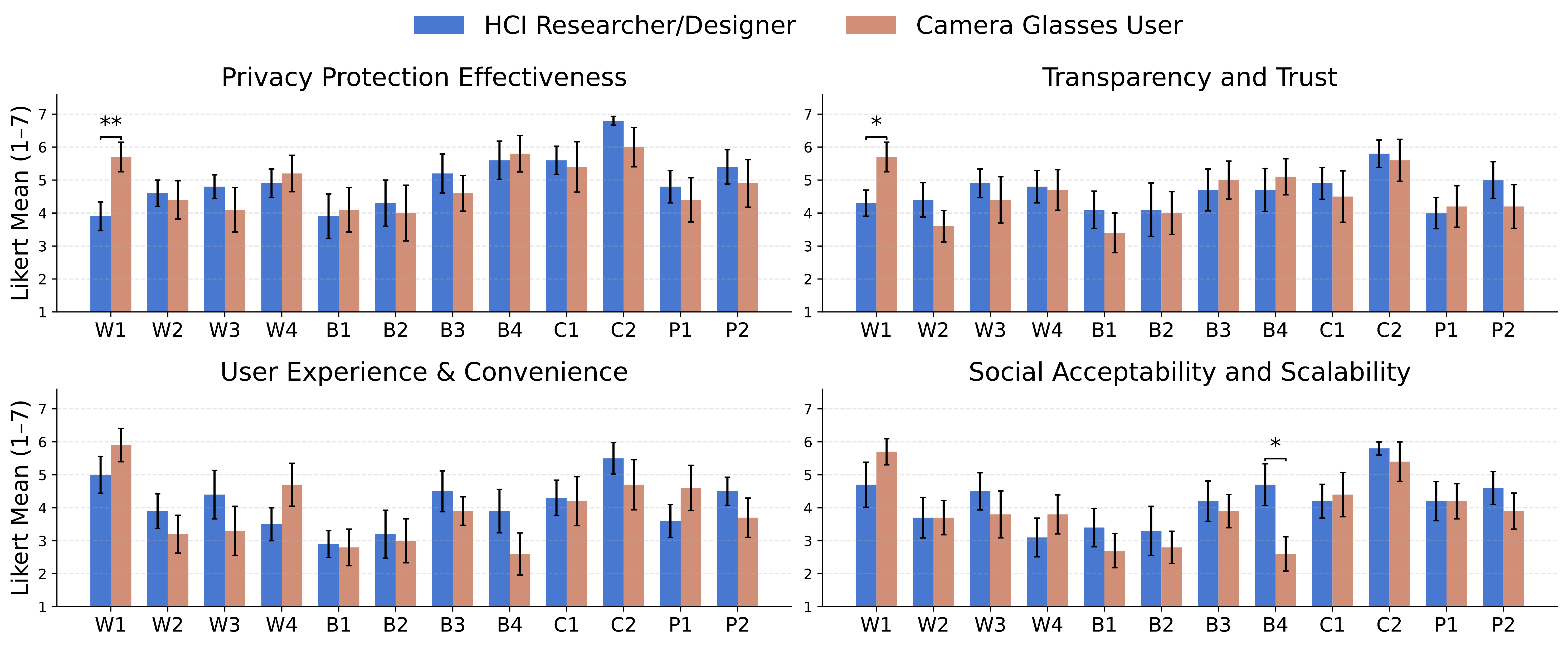}
\caption{Comparative evaluation of 12 PETs by HCI researchers/designers (blue) and camera glasses users (red) across four dimensions. Error bars represent standard error. Asterisks indicate significant differences between groups (* p<.05, ** p<.01).}
\label{fig:pet-ratings}
\end{figure*}

\subsection{Overview of PET Evaluation Results}

Quantitative ratings across four evaluation dimensions reveal distinct category-level performance patterns and systematic stakeholder differences (Figure~\ref{fig:pet-ratings}).

\textbf{Context-aware automatic processing (C1-C2)} achieved the highest ratings across nearly all dimensions. Geofencing Control (C2) received exceptional privacy protection scores (HCI: M=6.8, Users: M=6.0), substantially outperforming all other mechanisms. Face Anonymization (C1) also performed strongly (HCI: M=5.6, Users: M=5.4), though with moderate usability concerns.

\textbf{Wearer-side awareness mechanisms (W1-W4)} showed mixed reception. Users rated LED Ring Indicator (W1) favorably, while HCI researchers expressed skepticism about their effectiveness (M=3.9 vs. M=5.7, p<.01). Audio Alerts (W2) and External Display (W3) received consistently mediocre ratings.

\textbf{Bystander-side consent mechanisms (B1-B4)} revealed fundamental trade-offs. Negotiation Platform (B4) achieved strong privacy protection scores (HCI: M=5.6, Users: M=5.8) but received the lowest usability ratings (Users: M=2.6). Gesture Recognition (B1) and Wearable Markers (B2) faced both effectiveness and acceptability challenges.

\textbf{Platform-level accountability systems (P1-P2)} received moderate ratings, suggesting perception as complementary rather than primary solutions.

Both groups generally agreed on relative mechanism rankings, yet notable divergences emerged. Users exhibited greater confidence in LED indicators, which HCI researchers deemed insufficient. For transparency and trust, users consistently rated awareness mechanisms higher (W1: M=5.7 vs. M=4.3, p<.05). Social acceptability ratings diverged most for negotiation platforms (B4: M=4.7 vs. M=2.6, p<.05). These patterns suggest that automated context-aware systems provide the optimal protection-usability balance, while consent mechanisms face adoption barriers despite privacy benefits.

\subsection{Context as Primary Determinant of PET Selection (RQ3)}

\begin{figure*}[t]
\centering
\includegraphics[width=\textwidth]{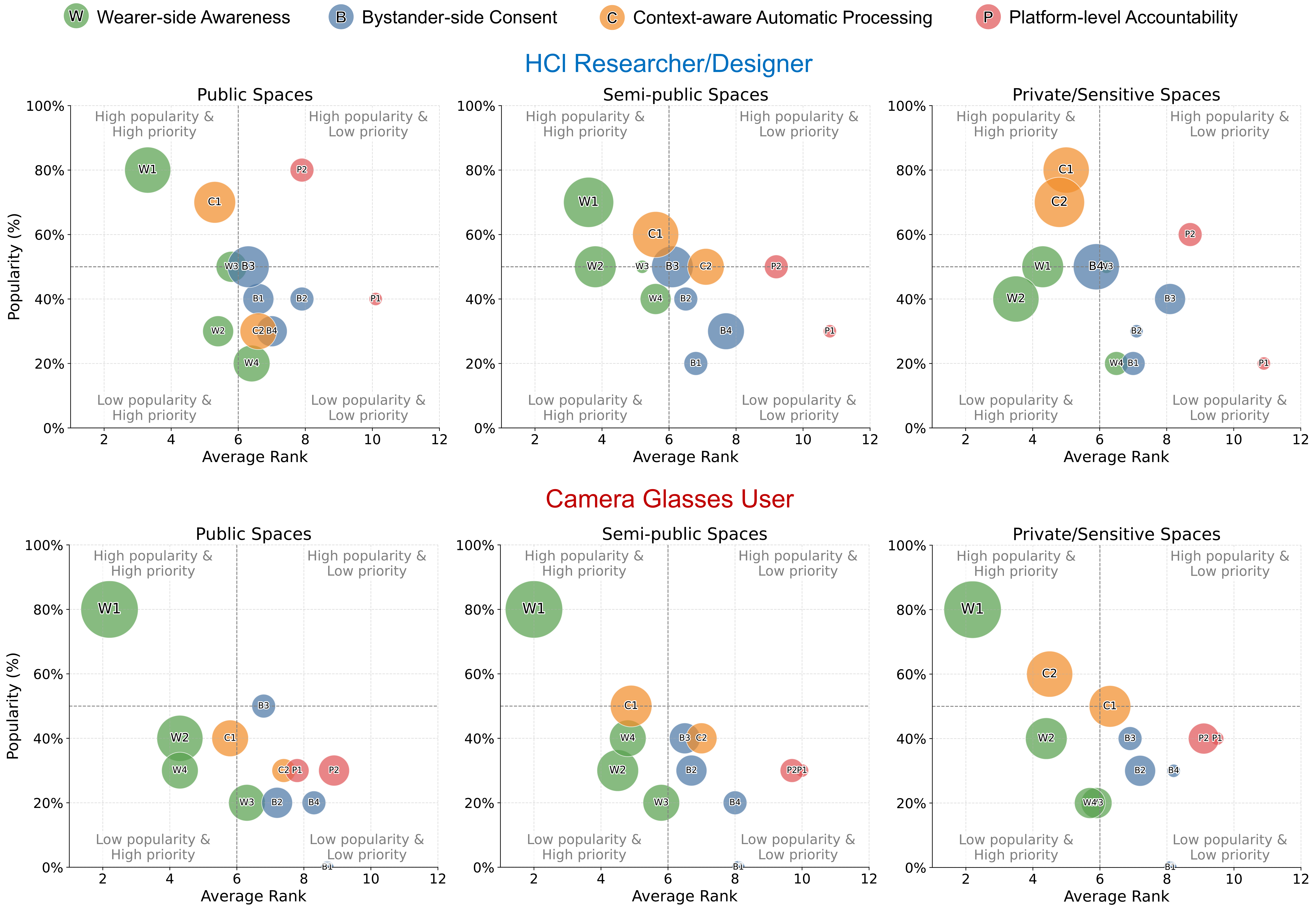}
\caption{Context-dependent privacy mechanism preferences across public, semi-public, and private spaces. Each point represents one mechanism positioned by average rank (x-axis) and selection popularity (y-axis). Bubble size indicates the proportion of participants ranking it in their top three.}
\label{fig:context-preferences}
\end{figure*}

Physical and social context emerged as the primary determinant of privacy mechanism preferences, overriding individual and role-based differences. Participants systematically prioritized different mechanism categories based on environmental characteristics (Figure~\ref{fig:context-preferences}).

\subsubsection{Public Spaces: Minimal-Friction Visibility}
Public environments generated remarkable convergence around passive visibility mechanisms. LED Ring Indicator (W1) achieved 80\% selection rates across both groups. This reflected practical scalability constraints: \textit{``In public places, visible indicators are sufficient. Other complex mechanisms have costs that are too high...there are just too many people''} (PH10). Interactive consent mechanisms faced systematic rejection, with gesture recognition receiving 0\% selection from users: \textit{``Actively requesting permission from everyone is just not practical''} (PU10). Face Anonymization (C1) emerged as the preferred complement (70\% HCI selection), addressing incidental capture through automatic rather than individual protection.

\subsubsection{Semi-Public Spaces: Structured Negotiation}
Professional and institutional environments created different preference patterns. Audio Alerts (W2) gained acceptance (50\% HCI selection): \textit{``In smaller scenarios involving personal privacy, voice announcements work for both parties''} (PH1). Negotiation Platform (B4) and Preference Broadcasting (B3) achieved higher support, reflecting the feasibility of structured consent in bounded environments where explicit communication is socially expected.

\subsubsection{Private/Sensitive Spaces: Automated Protection}
Private contexts revealed distinctions between trust-based gatherings and vulnerability-based sensitive spaces. For home gatherings, participants relied on social relationships: \textit{``Among friends, filming is no big deal...if someone has malicious intent, they're just not my friend anymore''} (PH2). However, sensitive spaces (gyms, changing rooms) triggered strong preferences for automated protection. Geofencing Control (C2) achieved its highest selection rates (60-70\%): \textit{``For situations involving body exposure, automatic shutdown is best, regardless of who's filming''} (PU2). This vulnerability-based reasoning prioritized comprehensive protection over user autonomy in high-stakes contexts.

\subsection{Fundamental Trade-offs in Current PETs (RQ4)}

Our evaluation revealed four fundamental trade-offs that undermine current PET approaches, explaining why no single mechanism category can reconcile the stakeholder conflicts identified in Study 1.

\subsubsection{Visibility vs. Disruption}
Awareness mechanisms (W1-W4) face an irreconcilable tension: mechanisms sufficiently noticeable to inform bystanders inevitably disrupt social interactions, while subtle approaches fail to provide meaningful transparency.

LED indicators exemplify this contradiction. Users viewed dynamic lighting as intuitive: \textit{``You immediately sense these are electronic glasses''} (PU10). However, researchers identified critical limitations: environmental dependency (imperceptible in bright daylight), lack of standardization across products, and circumvention vulnerability: \textit{``Just physically block it...two millimeters of tape, algorithms can't detect it''} (PH4). This ``deters the honest but not the malicious'' nature undermines protection purposes.

Audio Alerts (W2) and External Display (W3) attempted to address visibility limitations but introduced severe social friction. Audio notifications faced rejection: \textit{``This is so awkward...suddenly a loud voice during photography''} (PU10). External displays offered clarity but at aesthetic cost: \textit{``It affects appearance so much I wouldn't buy it''} (PU3). Proximity Broadcast (W4) achieved highest privacy protection scores (5.05/7) but lowest social acceptability (3.45/7), with industry insiders noting that phone manufacturers would block such functionality (PU9). Even successful implementation wouldn't solve the core issue: \textit{``What's the point of awareness? I know, but can I stop it?''} (PU4).

\subsubsection{Empowerment vs. Burden}
Consent mechanisms (B1-B4) suffer from a fundamental paradox: empowering bystanders requires burdening them with responsibilities that should belong to those creating privacy risks.

Participants identified the injustice of requiring potential victims to actively defend themselves. For Gesture Recognition (B1): \textit{``The responsibility for privacy invasion lies with the recorder, but now you're making the recorded party perform gestures...transferring the burden to the victim''} (PU8). This sentiment intensified with Wearable Markers (B2): \textit{``It's like victim-blaming...the fundamental problem lies with those creating danger''} (PH3).

App-based mechanisms (B3-B4) offered sophisticated control but with severe practical costs: platform fragmentation, battery drain, and exclusion of populations without smartphones. Negotiation Platform (B4) achieved strong privacy scores (5.8/7) but lowest usability ratings (2.6/7): \textit{``If I encounter 100 people at a tourist site, I'd have to communicate with all 100. The moment I wanted to capture would be gone''} (PU10). The enforcement dilemma proved critical: mandatory enforcement eliminates user discretion, while optional compliance becomes merely symbolic. Direct interpersonal communication consistently emerged as preferred: \textit{``Just tell the glasses wearer `don't record me,' isn't that better?''} (PU1).

\subsubsection{Protection vs. Agency}
Automated processing mechanisms (C1-C2) received highest privacy ratings yet create conflicts between protection goals and user autonomy.

Face Anonymization (C1) appealed conceptually but faced implementation challenges. Computational constraints proved significant: \textit{``Currently I can only record four 10-minute videos before battery death. Adding real-time blurring would reduce this further''} (PU6). The mechanism also cannot solve identification problems: \textit{``I might think I'm filming this handsome guy but actually have him off to the side while recording the pretty girl in the center. Who gets blurred?''} (PU9).

Geofencing Control (C2) achieved highest ratings overall (Privacy Protection: Users 6.0, HCI 6.8) but faces technical barriers: GPS cannot distinguish floors, WiFi connectivity is limited, and computational demands would reduce battery life dramatically. Philosophical disagreements emerged, with some participants viewing mandatory restrictions as \textit{``deliberately crippled products''} while others saw them as embedding legal compliance. Both mechanisms struggled with contextual nuance, as artistic and documentation needs clashed with blanket restrictions. A critical weakness emerged around transparency: \textit{``The person being recorded still doesn't know about processing...they would still feel uncomfortable''} (PH7). However, the high ratings for geofencing despite implementation challenges reveal acceptance of reduced functionality when privacy stakes are highest: \textit{``Some places absolutely shouldn't allow recording, like bathrooms, and this should apply universally''} (PH6).

\subsubsection{Accountability vs. Exposure}
Platform-level mechanisms (P1-P2) attempt deterrence through post-capture consequences but require surrendering privacy to protect it.

Digital Watermarking (P1) requires mandatory real-name authentication and device binding, creating surveillance infrastructure: \textit{``If this mechanism existed, I probably wouldn't buy the device. It violates the recorder's rights while protecting the recorded person''} (PU8). Face Matching (P2) exemplified this paradox more starkly: \textit{``You're building an extremely dangerous dataset to solve problems that may not even exist yet''} (PH2). Users must upload biometric data to platforms they don't trust, exposing themselves to risks potentially exceeding those they seek to avoid.

Both mechanisms faced scalability barriers and neither prevents initial violations, only offering potential post-hoc recourse. The moderate ratings (3.9-5.4) reflect recognition of deterrence value tempered by severe concerns about privacy surrenders and platform dependencies required for implementation.

\section{Context-Adaptive Privacy Pathways for Camera Glasses}

Through two complementary studies examining privacy negotiation from multiple stakeholder perspectives across diverse contexts, our work offers a systematic account of how privacy expectations diverge and how PETs succeed or fail under different conditions. In this section, we translate these insights into context-adaptive design framework and discuss potential pathways for future PET development. We emphasize that these pathways represent a proposed design framework grounded in our empirical findings rather than a validated system implementation.

\begin{figure*}[t]
\centering
\includegraphics[width=\textwidth]{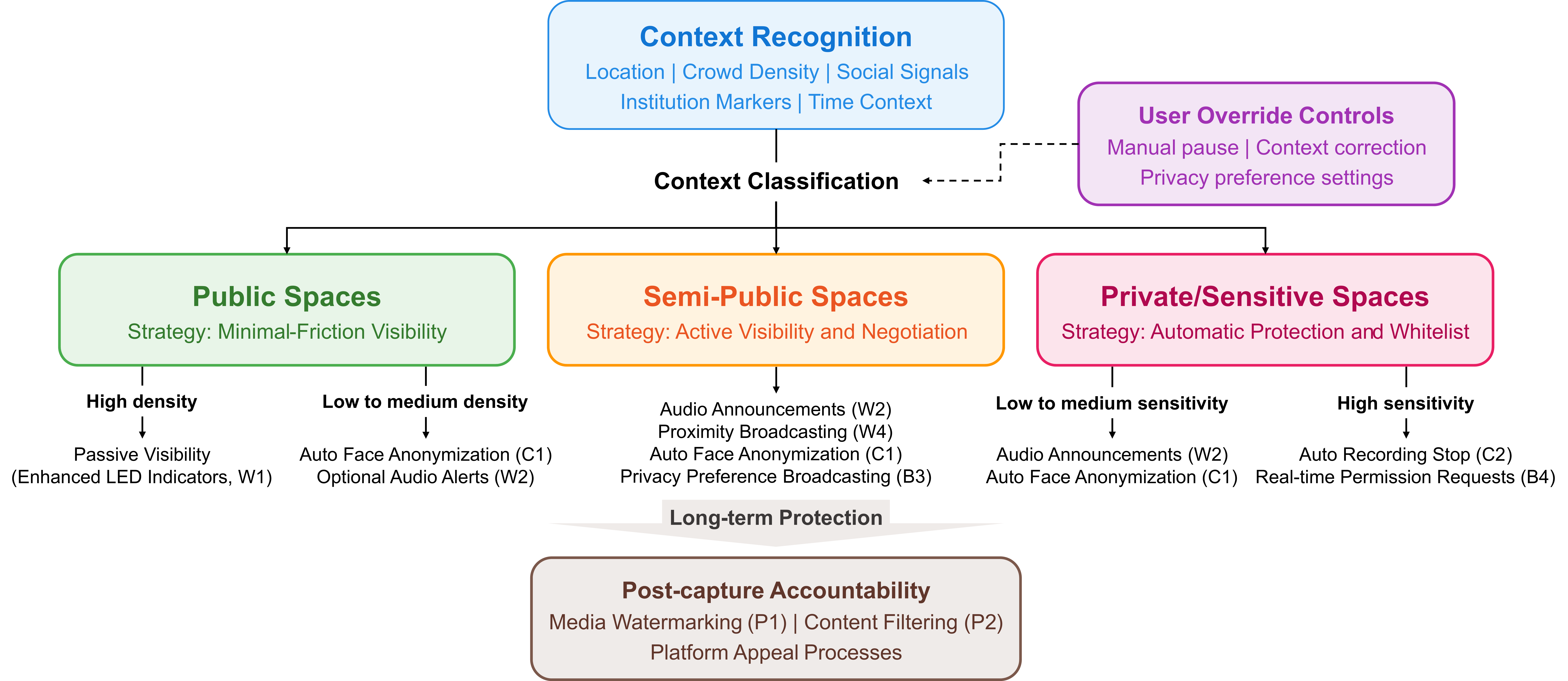}
\caption{Proposed context-adaptive privacy protection pathways for camera glasses. The system recognizes contextual characteristics, classifies environments into primary categories, and deploys appropriate mechanism combinations while preserving user override controls.}
\label{fig:adaptive-architecture}
\end{figure*}

\subsection{Key Patterns for Context-Adaptive Design}

Our studies reveal three key patterns that inform context-adaptive design.

\textbf{Asymmetric contextual sensitivity.} Wearers and bystanders exhibit fundamentally different relationships with context. Wearers maintained relatively stable recording reasonability perceptions across all scenarios (means $>$ 4.0), treating context as modulating the \textit{degree} of acceptable recording. Bystanders showed dramatic contextual variation ($F=56.440$, $p<.001$), with sensitive contexts generating nearly 2-point higher concerns than public settings. This asymmetry reveals a fundamental disconnect: wearers view context as adjusting ``how much'' disclosure is needed, while bystanders experience context as determining ``whether'' recording should occur at all.

\textbf{Gap concentration in control dimensions.} Expectation-willingness gaps systematically amplify in scenarios characterized by vulnerability. Gym $\times$ Strangers produced the largest disparities across data sharing ($\Delta M=1.04$), prior consent ($\Delta M=0.81$), and content transparency ($\Delta M=0.98$). Critically, these gaps concentrate in control dimensions (consent, data sharing) rather than transparency dimensions (purpose disclosure: $\Delta M=0.58$). This suggests that awareness-centric approaches, which dominate current designs, address secondary rather than primary privacy needs. As one participant noted: \textit{``What's the point of awareness? I know, but can I stop it?''} (PU4).

\textbf{Context-specific mechanism acceptance.} Study 2 revealed that stakeholder convergence emerged \textit{within} contexts but diverged \textit{across} contexts. Public spaces saw 80\% convergence on passive visibility, with interactive consent receiving 0\% user selection. Semi-public spaces enabled negotiation platforms that were rejected elsewhere. Sensitive spaces triggered strong preferences for automatic protection (60--70\% selection), with participants prioritizing comprehensive protection over user autonomy. These patterns indicate that effective privacy protection requires adaptive strategies calibrated to environmental and social characteristics rather than universal solutions.

\subsection{Proposed Context-Adaptive Pathways}

The fundamental trade-offs revealed across PET categories (visibility versus disruption, empowerment versus burden, protection versus agency, accountability versus exposure) demonstrate that static, universal mechanisms cannot achieve effective privacy protection. Based on our findings, we propose a context-adaptive framework operating on three core pathways (Figure~\ref{fig:adaptive-architecture}).

\textbf{Environmental recognition} identifies contextual characteristics through location data, crowd density estimation, institutional markers, and temporal patterns without invasive monitoring. \textbf{Dynamic strategy selection} deploys distinct protection approaches: public spaces activate minimal-friction visibility (W1) with optional face anonymization (C1); semi-public environments enable structured negotiation through audio announcements (W2) and preference broadcasting (B3); sensitive spaces trigger automatic recording restrictions (C2) or permission-based access (B4). \textbf{Layered protection} maintains baseline protections universally while activating context-triggered mechanisms and preserving user override capabilities~\cite{kudina2019ethics}.

These adaptive pathways addresses the fundamental trade-offs identified in our analysis. The visibility contradiction resolves through context-appropriate notification intensity. The control dilemma transforms from universal burden to selective empowerment, with bystanders gaining automatic protections in vulnerable contexts rather than requiring constant self-defense. The automation challenge becomes contextually bounded, preserving user agency in public spaces while accepting intervention where vulnerability justifies reduced control.

\subsection{Illustrative Application Scenarios}

To demonstrate the applications of context-adaptive pathways, we constructed three representative scenarios across public, semi-public, and sensitive environments. These scenarios track Information Transparency (IT) and Protective Measures (PM), the two dimensions where expectation-willingness gaps were most pronounced, illustrating how each pathway dynamically adjusts protection strategies in response to evolving contextual cues.

\subsubsection{Public Space: Vlogger in a Park}

\begin{figure*}[t]
    \centering
    \includegraphics[width=\textwidth]{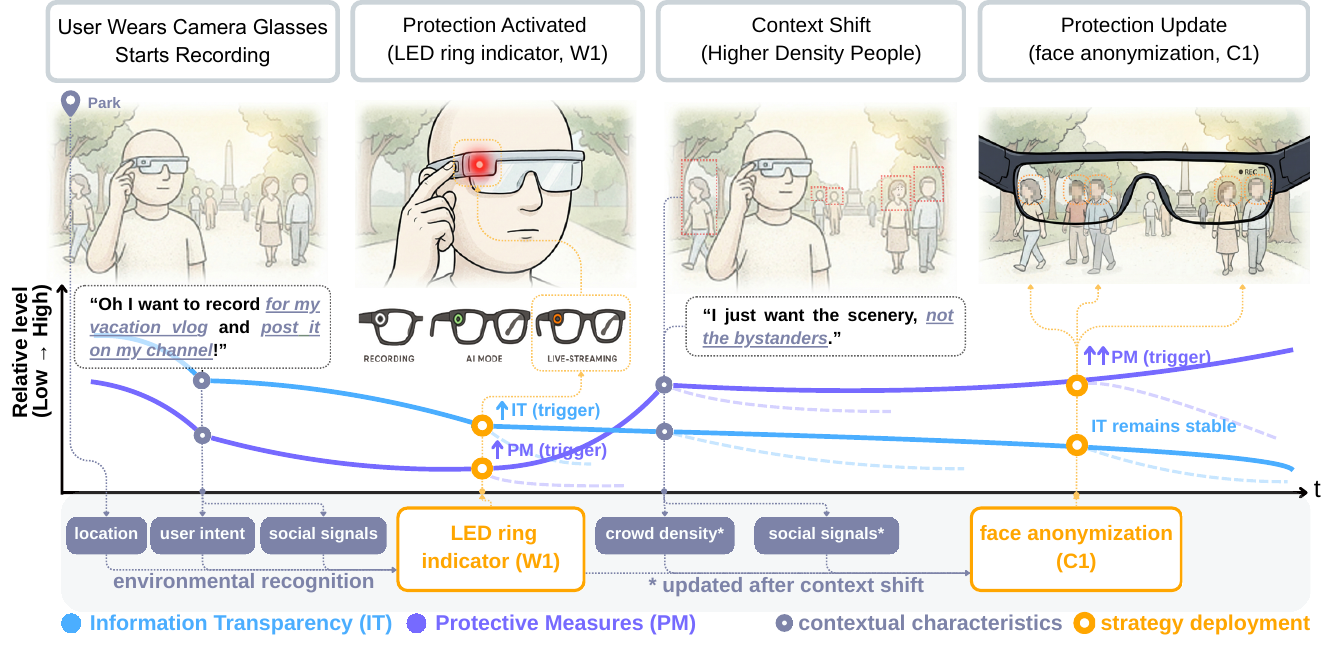}
    \caption{Context-adaptive privacy protection for a vlogger in a public park. The system adjusts strategies as contextual cues evolve: enhanced LED ring indicators (W1) stabilize IT when recording begins, and face anonymization (C1) raises PM as crowd density increases. Dashed curves show trajectories without intervention.}
    \label{fig:publicspaced}
    \vspace{5mm}
\end{figure*}

Figure~\ref{fig:publicspaced} illustrates a vlogger recording while jogging. When the wearer activates recording, both IT and PM decline as bystanders face uncertainty about capture status. The system detects recording activation and recommends enhanced LED ring indicators (W1) with semantically meaningful states (white for recording, green for AI processing, red for livestreaming), stabilizing IT and modestly improving PM. As crowd density increases and more identifiable faces appear, contextual recognition updates its assessment and activates automatic face anonymization (C1), substantially raising PM while maintaining stable IT. Without these interventions (dashed curves), both metrics would continue deteriorating.

\begin{figure*}[t]
    \centering
    \includegraphics[width=\textwidth]{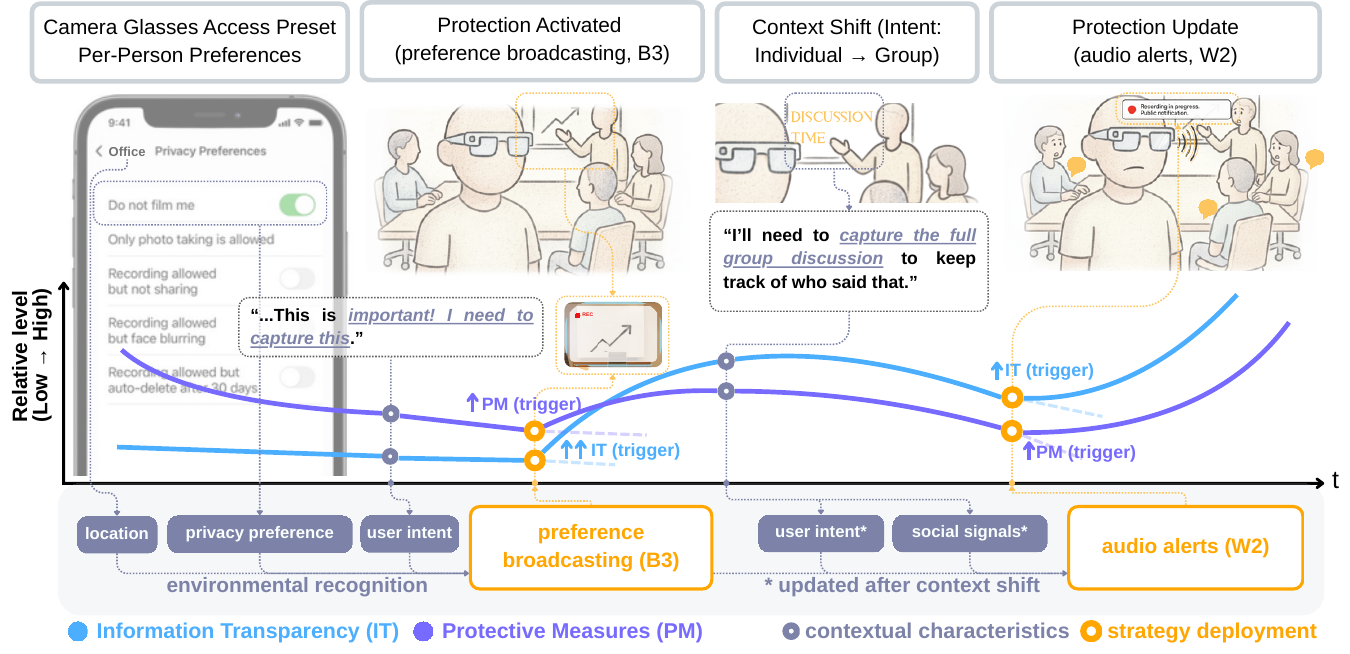}
    \caption{Context-adaptive privacy protection during an office meeting. Preset preferences enable proximity broadcasting (B3); explicit consent requests trigger audio alert (W2) when recording scope expands.}
    \label{fig:semi}
\end{figure*}

\begin{figure*}[t]
    \centering
    \includegraphics[width=\textwidth]{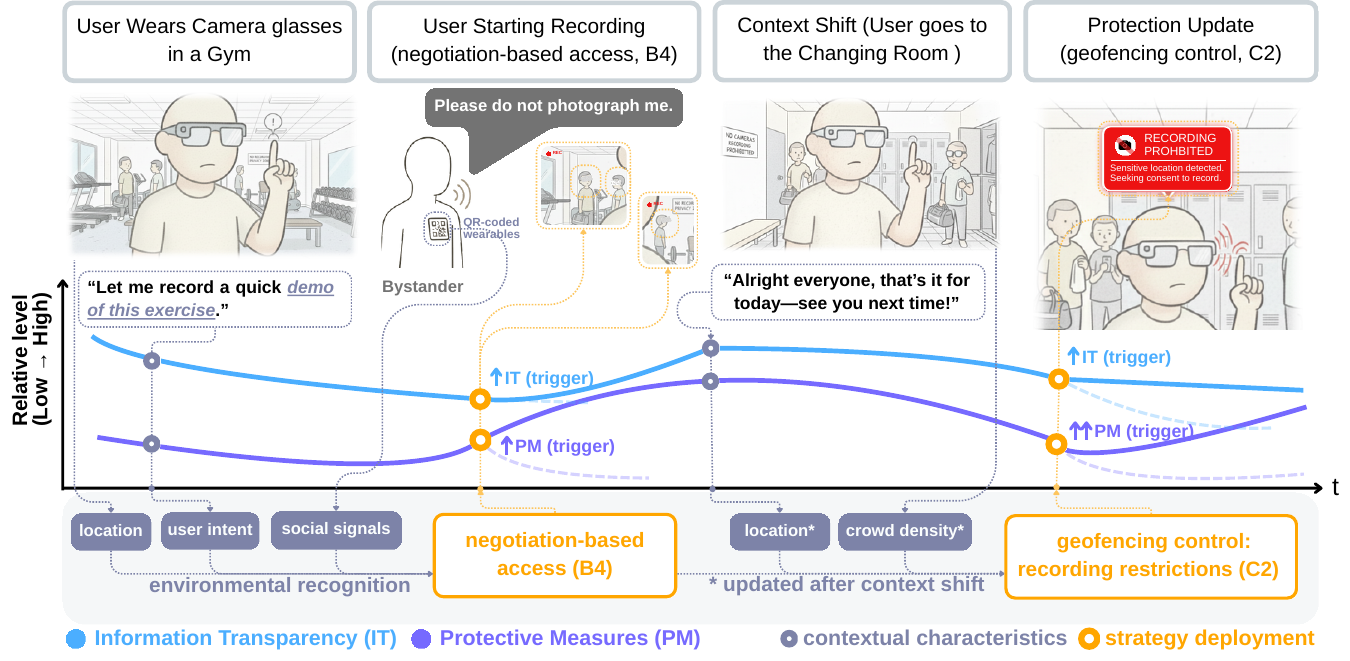}
    \caption{Context-adaptive privacy protection in a gym. Face anonymization via negotiation-based access (B4) activates when opt-out markers are detected; automatic geofencing control (C2) triggers upon entering the changing room.}
    \label{fig:private}
\end{figure*}

\subsubsection{Semi-Public Space: Office Meeting}

Figure~\ref{fig:semi} depicts an employee capturing meeting notes. Colleagues share a virtual workspace where they pre-configure privacy preferences (e.g., ``Do not film me,'' ``Photo only''). These preferences are automatically broadcast to nearby devices as the meeting begins.

When the wearer decides to capture the whiteboard, the system triggers proximity broadcasting (B3) based on the preset preference as the recommended protection. Since this aligns with default permissions, no explicit negotiation is required, and IT and PM stabilize. Later, when the wearer requests to capture the group discussion itself, contextual recognition detects this intent shift and activates consent-based audio alerts (W2). IT rises as everyone receives clear notification, while PM increases as colleagues can adjust their positions or explicitly consent.

\subsubsection{Sensitive Space: Gym and Changing Room}

Figure~\ref{fig:private} follows a fitness instructor recording an exercise demonstration. When recording begins, IT and PM decline as the action introduces potential incidental capture. As individuals with opt-out markers (e.g., QR-coded wearables) enter the frame, the system activates contextual face anonymization via negotiation-based access (B4), thereby stabilizing IT and raising PM.

When the instructor finishes the livestreaming sessions and enters the changing room, contextual recognition detects transition into a highly sensitive location. The system triggers geofencing control (C2) --- an automatic recording shutdown, producing a sharp PM increase as mandatory restrictions take effect in protected spaces. This scenario illustrates how adaptive systems can enforce graduated protection, from voluntary anonymization in general gym areas to mandatory cessation in spaces where vulnerability justifies overriding user autonomy.

Together, these scenarios illustrate how context-adaptive pathways can dynamically mediate the expectation-willingness gap by selecting appropriate protection strategies based on environmental characteristics, user intent, and bystander signals.

\subsection{Implementation Challenges and Future Directions}

Realizing context-adaptive privacy requires coordination across technical, regulatory, and social dimensions. As one participant noted, \textit{``Many issues aren't things a single manufacturer can solve—they involve upstream and downstream platforms, government policies, and user education''} (PH10).

\textbf{Near-term enhancements} should focus on improving current mechanisms. Passive indicators (W1) will likely remain the most accepted notification form but must evolve beyond theatrical compliance. Transparent hardware design should replace concealed recording~\cite{koelle2018beyond}, and manual camera covers can provide tangible trust signals~\cite{do2021smart}. Multi-modal feedback (enhanced LED visibility, selective audio cues, haptic confirmation) can address situational constraints while minimizing disruption~\cite{wu2024designing, ahmad2020tangible}. Hybrid approaches combining user input with automated detection may address the algorithmic difficulty of distinguishing subjects from bystanders~\cite{niu2025not, steil2019privaceye, windl2025designing}.

\textbf{Medium-term development} should pursue context-adaptive systems. Binary recording approaches should give way to graduated protection levels~\cite{cruz2025assessing}: original capture for trusted contexts, aesthetic-preserving filters for public spaces, complete restriction for sensitive areas. Such adaptive defaults are critical since users rarely modify initial configurations~\cite{abraham2024don}. Location, social density, and institutional cues can inform automatic adjustments, while contextual preference prediction~\cite{yang2024feasibility} may reduce user burden. Lightweight negotiation protocols prioritizing immediacy, such as gesture-based responses~\cite{barhm2011negotiating} or pre-configured permissions~\cite{aditya2016pic, shu2016cardea}, can support ephemeral encounters.

\textbf{Long-term ecosystem development} requires industry standardization elevating privacy protection from user discretion to system-level enforcement~\cite{venugopalan2024aragorn}, specifying minimum notification requirements and cross-manufacturer interoperability. Regulatory frameworks must address ubiquitous sensing devices specifically, moving beyond traditional consent models toward environmental protection standards. Social norm development must complement technical measures~\cite{bhardwaj2024focus, o2023augmenting}, with public education establishing expectations for voluntary restraint in sensitive contexts and organizations developing clear camera-glasses policies.

These efforts collectively mark a shift from binary privacy approaches toward contextually intelligent systems that recognize privacy as dynamically shaped by environmental and social characteristics.


\section{Discussion}

\subsection{Contextual Asymmetry in Privacy Negotiation}
\label{6.1}

Context emerged as the primary determinant of privacy acceptability in both studies. Study 1 showed that physical settings and social relationships intensify the expectation-willingness gap, highlighting context as a fluid, co-constructed frame that shapes how people interpret capture~\cite{kudina2019ethics, windl2025designing}. Because camera-glasses privacy centers on whether sensing should occur rather than how information flows afterward, our findings extend classic privacy models from static information management toward ongoing contextual negotiation~\cite{altman1975environment, nissenbaum2004privacy}. Study 2 further demonstrates that PET preferences shift with these contextual interpretations, underscoring the need for adaptive pathways that accommodate evolving notions of context and agency.

Our findings also surface additional contextual layers that influence capture judgments. Behavioral cues such as gesture, intent, timing, and framing~\cite{shu2018cardea, koelle2018your, gallardo2023speculative} actively reshape how a setting is perceived. For example, the same classroom may feel public or private depending on what is being recorded and the wearer's visible behavior. Interface-level cues, including visual changes or capture-mode indicators~\cite{windl2025designing, rajaram2023reframe}, can guide how people make sense of sensing in the moment. Personal differences including cultural norms~\cite{ghaiumy2021difficulties, roberts2017privacy}, accessibility needs~\cite{zhao2023if, ahmed2018up}, and individual traits~\cite{egelman2015predicting} create further opportunities for systems to refine contextual awareness and provide more equitable protection.

Rather than exhaustively modeling every contextual factor, our contribution lies in revealing how these elements gain significance in lived experience. By grounding privacy negotiation in situated user interpretation rather than optimization metrics~\cite{rajaram2025privacy}, we point toward adaptive mechanisms that evolve with social interaction and contextual meaning.

\subsection{Device Familiarity and Social Acceptance}
\label{6.2}

Although familiarity was not the dominant determinant of privacy judgments, it shaped how participants evaluated camera-glasses acceptability. Increasing familiarity was associated with reduced privacy concern and lower demands for transparency, consistent with technology-familiarity and social-acceptance models~\cite{carroll2002adoption, denning2014situ}. Professional settings amplified this effect, where normative expectations of documentation generated substantially higher passive acceptance (35\%) than other contexts (10--15\%). These patterns suggest that normalization can soften perceived risk.

Yet increasing familiarity did not eliminate wearer-bystander divergence. The expectation-willingness gap remained significant even at the ``very familiar'' level. Wearers showed a sharp drop in concern, whereas bystanders' decline was gradual, suggesting that deeper understanding does not guarantee convergence but instead reveals enduring asymmetry between roles.

A notable reversal emerged in evaluations of LED indicator sufficiency. For users with low familiarity, ``the light is on'' signifies transparency and safety, with acceptance peaking at the ``very familiar'' level. However, current/former users recognize that LEDs are visible yet socially ineffective~\cite{thakkar2022would, ahmad2020tangible}. This \textit{familiarity paradox}~\cite{koelle2018beyond} suggests that as users gain deeper understanding of camera-glasses operation, familiarity no longer promotes acceptance but instead reveals the device's covert potential. This effect may be particularly salient for camera glasses, where capture is inherently less visible than in traditional cameras or HMDs~\cite{gallardo2023speculative}.

Interestingly, no comparable fluctuation appeared in information transparency or protection demands. One explanation is ``privacy resignation'': as familiarity increases, bystanders become accustomed to potential risks and pragmatically tolerate data vulnerability while feeling powerless to resist~\cite{marky2020don, draper2017privacy, bhardwaj2024focus}. Alternatively, expectations for transparency and protection may operate at a broader social level rather than being device-specific, producing stable ratings despite continued critical awareness~\cite{nissenbaum2004privacy, egelman2016behavior}. Future longitudinal work could examine how familiarity interacts with psychological adaptation and regulatory expectations over time.

\subsection{Scaling Privacy Negotiation to Complex Settings}
\label{6.3}

While our studies focused on dyadic interactions between a wearer and a bystander, real-world public spaces often involve dynamic, multi-party environments where traditional PETs become impractical due to interaction time and attentional costs~\cite{shu2017your, bo2014privacy}. Excessive notifications may intensify usability tensions as audiences grow.

The four trade-offs identified in Study 2 become more pronounced in multi-party settings. The visibility-disruption tension grows as collective awareness requires signals that quickly exceed social comfort. The empowerment-burden tension scales with group size, making multi-party consent unmanageable. The protection-agency tension persists because automated controls cannot eliminate the need for user autonomy. The accountability-exposure tension intensifies as responsibility extends across more actors. These amplified pressures indicate that multi-party privacy negotiation is not simply a scaled-up dyadic problem but a fundamentally different design space.

Building on this logic, adaptive privacy protocols could address these trade-offs by adjusting signals to reduce visibility-disruption tension, simplifying how bystanders express preferences to ease empowerment-burden tensions, and using context-aware automation that preserves user agency. Multi-modal cues may help clarify accountability without creating additional exposure~\cite{thakkar2022would, ahmad2020tangible}. Such approaches frame privacy not as individual negotiation but as an in-situ, collective process~\cite{kudina2019ethics}.

\subsection{Multi-Stakeholder Perspectives and Cultural Considerations}
\label{6.4}

Incorporating a multi-stakeholder perspective revealed that privacy expectations are role-dependent and context-sensitive~\cite{o2023privacy, denning2014situ}. Unlike prior work that mainly documents risks or evaluates isolated mechanisms~\cite{mansour2023bans, ahmad2020tangible}, our paired design directly captured experiences and demands from both groups within shared scenarios. This enabled systematic examination of how contextual features shape privacy judgments and helped identify key trade-offs in PET design~\cite{windl2025designing, corbett2023securing, chung2023negotiating}. In dynamic environments where individuals may shift roles constantly~\cite{bhardwaj2024focus}, understanding both perspectives provides a foundation for developing protection systems that respond to changing social configurations.

Cultural background further shapes these dynamics~\cite{ghaiumy2021difficulties, roberts2017privacy, li2022cultural}. Our Chinese participants, drawn from the world's largest camera glasses market ~\cite{xromXiaomiRaises, xromForecasts107}, showed high familiarity with domestic brands (87--89\% knew Xiaomi) compared to international products (Ray-Ban Meta: 27--44\%; Google Glass: 30--32\%), partially grounding their evaluations in locally dominant products. Furthermore, people in collectivist cultures may treat privacy more as a social matter, placing greater emphasis on interactional risks than individuals in more individualist contexts~\cite{li2022cultural}. Our sample shows this pattern: dense social networks (74\% knew users, 25\% knew 3+ users) coexist with strong control expectations (M = 5.99), indicating privacy boundaries shaped through social relationships rather than individual autonomy alone.

These findings suggest that privacy should be understood as a culturally situated negotiation rather than a universal preference. Complementing this work with studies from other societies can help understand how cultural norms~\cite{bhardwaj2024focus, kim2025cultural} influence privacy needs in shared environments. Cultural variation also intersects with identities such as age and ability~\cite{rao2020types, akter2020uncomfortable, zhao2023if}, requiring careful consideration when generalizing findings across contexts.

\section{Limitations and Future Work}

Several methodological limitations constrain the generalizability of our findings. Our exclusive focus on Chinese participants limits cross-cultural applicability. China's distinct privacy culture, rapid smart glasses adoption, and regulatory environment may produce patterns that do not generalize to Western contexts where privacy expectations differ substantially~\cite{miltgen2014cultural}. The timing of data collection during China's smart glasses market expansion may capture attitudes specific to early adoption phases rather than mature market dynamics.

Study 1's reliance on vignette-based scenarios, while enabling systematic manipulation, cannot capture the emotional intensity and social complexity of actual privacy violations~\cite{denning2014situ}. Self-reported measures introduce potential social desirability effects~\cite{fisher1993social} and privacy paradox concerns~\cite{kokolakis2017privacy}, though we attempted mitigation through indirect questioning, randomized scenario presentation, and inclusion of both attitudinal and behavioral intention measures. Additionally, our abbreviated baseline attitude scales (2 items each) showed modest internal consistency, particularly for Information Sharing Intention ($\alpha$ = 0.48). While convergent validity analyses supported their use as descriptive measures, future research employing baseline attitudes as primary variables should use longer, fully validated scales such as the VOPP~\cite{hasan2023psychometric} or complete IUIPC instruments~\cite{malhotra2004internet}.

Study 2's paired interviews may have encouraged consensus-seeking, and scenario-based PET evaluation cannot replicate real-world implementation constraints. Both samples skewed toward younger, educated, technology-literate participants, potentially underrepresenting broader attitudes. Systematic research on excluded groups, including elderly individuals, children, and people with disabilities, is essential for equitable protection. Importantly, our studies capture stated preferences rather than behavioral evidence. While we provide diagnostic value by quantifying gaps and identifying trade-offs, validating whether these preferences predict real-world behavior requires longitudinal field deployments.

Looking forward, comparative studies across Western and Eastern contexts are essential for developing globally applicable privacy frameworks. Technical research priorities include developing lightweight context recognition systems that identify environmental characteristics without surveillance~\cite{yang2024feasibility, rajaram2025exploring}, privacy-preserving negotiation protocols for rapid, anonymous preference expression in ephemeral encounters~\cite{alshehri2023exploring}, and aesthetic-preserving privacy filters that protect bystanders without destroying creative intent~\cite{cruz2025assessing, mujirishvili2024don}. Only through such comprehensive approaches can we enable the benefits of wearable cameras while preserving the privacy foundations essential for social trust.

\section{Conclusion}

This investigation provides a systematic multi-stakeholder evaluation of privacy mechanisms for camera glasses. Through surveys (N=525) and paired interviews (N=20) evaluating 12 PETs, we identified persistent expectation-willingness gaps stemming from four fundamental trade-offs: visibility versus disruption, empowerment versus burden, protection versus agency, and accountability versus exposure.

Context emerged as the primary determinant of privacy acceptability, extending contextual integrity theory to questions of whether sensing should occur. Our context-adaptive pathways progress from minimal-friction visibility in public spaces through structured negotiation in semi-public environments to automatic protection in sensitive contexts. These findings challenge assumptions that better technical mechanisms alone can reconcile privacy conflicts, pointing toward privacy as collectively determined and structurally enforced through coordinated sociotechnical systems.

\begin{acks}
This work is supported by the Natural Science Foundation of China (NSFC) under Grant No. 62472243. 
\end{acks}

\bibliographystyle{ACM-Reference-Format}
\bibliography{References}

@String{Computing = "Computing" }

@String{Computer = "{IEEE} Computer" }

@String{Springer = "Springer-Verlag" }

@misc{vmagazineRayBanMeta,
	author = {{V}.{M}agazine.},
	title = {{R}ay-{B}an {M}eta {S}mart {G}lasses {G}lobal {S}ales {S}urpass 2 {M}illion {U}nits! {T}he {N}ext {A}nticipated {T}arget {C}ould {B}e {P}rada {M}eta!},
	howpublished = {\url{https://www.vmagazine.hk/2025/07/29/ray-ban-meta-smart-glasses-global-sales-surpass-2-million-units-the-next-anticipated-target-could-be-prada-meta/}},
	year = {2025},
	note = {[Accessed 16-08-2025]},
}

@article{corbett2023securing,
  title={Securing bystander privacy in mixed reality while protecting the user experience},
  author={Corbett, Matthew and David-John, Brendan and Shang, Jiacheng and Hu, Y Charlie and Ji, Bo},
  journal={IEEE Security \& Privacy},
  volume={22},
  number={1},
  pages={33--42},
  year={2023},
  publisher={IEEE}
}

@article{o2023privacy,
  title={Privacy-enhancing technology and everyday augmented reality: Understanding bystanders' varying needs for awareness and consent},
  author={O'Hagan, Joseph and Saeghe, Pejman and Gugenheimer, Jan and Medeiros, Daniel and Marky, Karola and Khamis, Mohamed and McGill, Mark},
  journal={Proceedings of the ACM on Interactive, Mobile, Wearable and Ubiquitous Technologies},
  volume={6},
  number={4},
  pages={1--35},
  year={2023},
  publisher={ACM New York, NY, USA}
}

@inproceedings{denning2014situ,
  title={In situ with bystanders of augmented reality glasses: Perspectives on recording and privacy-mediating technologies},
  author={Denning, Tamara and Dehlawi, Zakariya and Kohno, Tadayoshi},
  booktitle={Proceedings of the SIGCHI conference on human factors in computing systems},
  pages={2377--2386},
  year={2014}
}

@misc{brookingsSeeingPast,
	author = {Wheeler, Tom},
	title = {{S}eeing past the cool: {F}acebook’s new smart glasses | {B}rooking.},
	howpublished = {\url{https://www.brookings.edu/articles/seeing-past-the-cool-facebooks-new-smart-glasses/}},
	year = {2021},
	note = {[Accessed 16-08-2025]},
}

@misc{techcrunchFacebookWarned,
	author = {Natasha Lomas},
	title = {{F}acebook warned over 'very small' indicator {L}{E}{D} on smart glasses, as {E}{U} {D}{P}{A}s flag privacy concerns.},
	howpublished = {\url{https://techcrunch.com/2021/09/20/facebook-warned-over-very-small-indicator-led-on-smart-glasses-as-eu-dpas-flag-privacy-concerns}},
	year = {2021},
	note = {[Accessed 16-08-2025]},
}

@inproceedings{bhardwaj2024focus,
  title={In Focus, Out of Privacy: The Wearer's Perspective on the Privacy Dilemma of Camera Glasses},
  author={Bhardwaj, Divyanshu and Ponticello, Alexander and Tomar, Shreya and Dabrowski, Adrian and Krombholz, Katharina},
  booktitle={Proceedings of the 2024 CHI Conference on Human Factors in Computing Systems},
  pages={1--18},
  year={2024}
}

@article{niu2025everyone,
  title={Everyone's Privacy Matters! An Analysis of Privacy Leakage from Real-World Facial Images on Twitter and Associated User Behaviors},
  author={Niu, Yuqi and Qiu, Weidong and Tang, Peng and Wang, Lifan and Chen, Shuo and Li, Shujun and K{\"o}kciyan, Nadin and Niu, Ben},
  journal={Proceedings of the ACM on Human-Computer Interaction},
  volume={9},
  number={2},
  pages={1--38},
  year={2025},
  publisher={ACM New York, NY, USA}
}

@inproceedings{lebeck2018towards,
  title={Towards security and privacy for multi-user augmented reality: Foundations with end users},
  author={Lebeck, Kiron and Ruth, Kimberly and Kohno, Tadayoshi and Roesner, Franziska},
  booktitle={2018 IEEE Symposium on Security and Privacy (SP)},
  pages={392--408},
  year={2018},
  organization={IEEE}
}

@inproceedings{o2023augmenting,
  title={Augmenting people, places \& media: The societal harms posed by everyday augmented reality, and the case for perceptual human rights},
  author={O'Hagan, Joseph and Gugenheimer, Jan and Bonner, Jolie and Mathis, Florian and McGill, Mark},
  booktitle={Proceedings of the 22nd International Conference on Mobile and Ubiquitous Multimedia},
  pages={225--235},
  year={2023}
}

@inproceedings{niu2025not,
  title={``I am not the primary focus"-Understanding the Perspectives of Bystanders in Photos Shared Online},
  author={Niu, Yuqi and Meng-Schneider, Nicole and Qiu, Weidong and Kokciyan, Nadin},
  booktitle={Proceedings of the 2025 CHI Conference on Human Factors in Computing Systems},
  pages={1--23},
  year={2025}
}

@article{de2019security,
  title={Security and privacy approaches in mixed reality: A literature survey},
  author={De Guzman, Jaybie A and Thilakarathna, Kanchana and Seneviratne, Aruna},
  journal={ACM Computing Surveys (CSUR)},
  volume={52},
  number={6},
  pages={1--37},
  year={2019},
  publisher={ACM New York, NY, USA}
}

@article{perez2017bystanders,
  title={Bystanders' privacy},
  author={Perez, Alfredo J and Zeadally, Sherali and Griffith, Scott},
  journal={IT Professional},
  volume={19},
  number={3},
  pages={61--65},
  year={2017},
  publisher={IEEE}
}

@inproceedings{david2024understanding,
  title={Understanding the long-term impact and perceptions of privacy-enhancing technologies for bystander obscuration in AR},
  author={David-John, Brendan and Ji, Bo and Selinger, Evan},
  booktitle={2024 IEEE International Symposium on Mixed and Augmented Reality Adjunct (ISMAR-Adjunct)},
  pages={23--25},
  year={2024},
  organization={IEEE}
}

@article{bukhari2025rethinking,
  title={Rethinking Privacy Indicators in Extended Reality: Multimodal Design for Situationally Impaired Bystanders},
  author={Bukhari, Syed Ibrahim Mustafa Shah and Sajid, Maha and Ji, Bo and David-John, Brendan},
  journal={arXiv preprint arXiv:2508.07057},
  year={2025}
}

@inproceedings{corbett2023bystandar,
  title={Bystandar: Protecting bystander visual data in augmented reality systems},
  author={Corbett, Matthew and David-John, Brendan and Shang, Jiacheng and Hu, Y Charlie and Ji, Bo},
  booktitle={Proceedings of the 21st Annual International Conference on Mobile Systems, Applications and Services},
  pages={370--382},
  year={2023}
}

@inproceedings{hasan2020automatically,
  title={Automatically detecting bystanders in photos to reduce privacy risks},
  author={Hasan, Rakibul and Crandall, David and Fritz, Mario and Kapadia, Apu},
  booktitle={2020 IEEE Symposium on Security and Privacy (SP)},
  pages={318--335},
  year={2020},
  organization={IEEE}
}

@inproceedings{koelle2018beyond,
  title={Beyond LED status lights-design requirements of privacy notices for body-worn cameras},
  author={Koelle, Marion and Wolf, Katrin and Boll, Susanne},
  booktitle={Proceedings of the Twelfth International Conference on Tangible, Embedded, and Embodied Interaction},
  pages={177--187},
  year={2018}
}

@inproceedings{koelle2018your,
  title={Your smart glasses' camera bothers me! exploring opt-in and opt-out gestures for privacy mediation},
  author={Koelle, Marion and Ananthanarayan, Swamy and Czupalla, Simon and Heuten, Wilko and Boll, Susanne},
  booktitle={Proceedings of the 10th Nordic Conference on Human-Computer Interaction},
  pages={473--481},
  year={2018}
}

@inproceedings{shu2017your,
  title={Your privacy is in your hand: Interactive visual privacy control with tags and gestures},
  author={Shu, Jiayu and Zheng, Rui and Hui, Pan},
  booktitle={International Conference on Communication Systems and Networks},
  pages={24--43},
  year={2017},
  organization={Springer}
}

@article{perez2018facepet,
  title={FacePET: Enhancing bystanders’ facial privacy with smart wearables/internet of things},
  author={Perez, Alfredo J and Zeadally, Sherali and Matos Garcia, Luis Y and Mouloud, Jaouad A and Griffith, Scott},
  journal={Electronics},
  volume={7},
  number={12},
  pages={379},
  year={2018},
  publisher={MDPI}
}

@article{shu2016cardea,
  title={Cardea: Context-aware visual privacy protection from pervasive cameras},
  author={Shu, Jiayu and Zheng, Rui and Hui, Pan},
  journal={arXiv preprint arXiv:1610.00889},
  year={2016}
}

@inproceedings{pidcock2011notisense,
  title={Notisense: An urban sensing notification system to improve bystander privacy},
  author={Pidcock, Sarah and Smits, Rob and Hengartner, Urs and Goldberg, Ian},
  booktitle={Proceedings of the 2nd International Workshop on Sensing Applications on Mobile Phones (PhoneSense), Seattle, WA, USA},
  pages={12--15},
  year={2011}
}

@inproceedings{aditya2016pic,
  title={I-pic: A platform for privacy-compliant image capture},
  author={Aditya, Paarijaat and Sen, Rijurekha and Druschel, Peter and Joon Oh, Seong and Benenson, Rodrigo and Fritz, Mario and Schiele, Bernt and Bhattacharjee, Bobby and Wu, Tong Tong},
  booktitle={Proceedings of the 14th annual international conference on mobile systems, applications, and services},
  pages={235--248},
  year={2016}
}

@inproceedings{li2016privacycamera,
  title={Privacycamera: Cooperative privacy-aware photographing with mobile phones},
  author={Li, Ang and Li, Qinghua and Gao, Wei},
  booktitle={2016 13th Annual IEEE International Conference on Sensing, Communication, and Networking (SECON)},
  pages={1--9},
  year={2016},
  organization={IEEE}
}

@inproceedings{sun2020iryp,
  title={iRyP: a purely edge-based visual privacy-respecting system for mobile cameras},
  author={Sun, Yuanyi and Chen, Shiqing and Zhu, Sencun and Chen, Yu},
  booktitle={Proceedings of the 13th ACM conference on security and privacy in wireless and mobile networks},
  pages={195--206},
  year={2020}
}

@article{perez2020user,
  title={A user study of a wearable system to enhance bystanders’ facial privacy},
  author={Perez, Alfredo J and Zeadally, Sherali and Griffith, Scott and Garcia, Luis Y Matos and Mouloud, Jaouad A},
  journal={IoT},
  volume={1},
  number={2},
  pages={13},
  year={2020},
  publisher={MDPI}
}

@inproceedings{pierce2022addressing,
  title={Addressing adjacent actor privacy: Designing for bystanders, co-users, and surveilled subjects of smart home cameras},
  author={Pierce, James and Weizenegger, Claire and Nandi, Parag and Agarwal, Isha and Gram, Gwenna and Hurrle, Jade and Liao, Hannah and Lo, Betty and Park, Aaron and Phan, Aivy and others},
  booktitle={Proceedings of the 2022 ACM Designing Interactive Systems Conference},
  pages={26--40},
  year={2022}
}

@inproceedings{mansour2023bans,
  title={Bans: Evaluation of bystander awareness notification systems for productivity in vr},
  author={Mansour, Shady and Knierim, Pascal and O’Hagan, Joseph and Alt, Florian and Mathis, Florian},
  booktitle={Network and Distributed Systems Security (NDSS) Symposium},
  volume={2},
  year={2023}
}

@article{ahmad2020tangible,
  title={Tangible privacy: Towards user-centric sensor designs for bystander privacy},
  author={Ahmad, Imtiaz and Farzan, Rosta and Kapadia, Apu and Lee, Adam J},
  journal={Proceedings of the ACM on Human-Computer Interaction},
  volume={4},
  number={CSCW2},
  pages={1--28},
  year={2020},
  publisher={ACM New York, NY, USA}
}

@inproceedings{wu2024designing,
  title={Designing the informing process with streamers and bystanders in live streaming},
  author={Wu, Yanlai and Gui, Xinning and Luo, Yuhan and Li, Yao},
  booktitle={Twentieth Symposium on Usable Privacy and Security (SOUPS 2024)},
  pages={315--332},
  year={2024}
}

@inproceedings{liao2019understanding,
  title={Understanding the role of privacy and trust in intelligent personal assistant adoption},
  author={Liao, Yuting and Vitak, Jessica and Kumar, Priya and Zimmer, Michael and Kritikos, Katherine},
  booktitle={International conference on information},
  pages={102--113},
  year={2019},
  organization={Springer}
}

@article{vimalkumar2021okay,
  title={‘Okay google, what about my privacy?’: User's privacy perceptions and acceptance of voice based digital assistants},
  author={Vimalkumar, M and Sharma, Sujeet Kumar and Singh, Jang Bahadur and Dwivedi, Yogesh K},
  journal={Computers in Human Behavior},
  volume={120},
  pages={106763},
  year={2021},
  publisher={Elsevier}
}

@inproceedings{rashidi2018you,
  title={" You don't want to be the next meme": College Students' Workarounds to Manage Privacy in the Era of Pervasive Photography},
  author={Rashidi, Yasmeen and Ahmed, Tousif and Patel, Felicia and Fath, Emily and Kapadia, Apu and Nippert-Eng, Christena and Su, Norman Makoto},
  booktitle={Fourteenth Symposium on Usable Privacy and Security (SOUPS 2018)},
  pages={143--157},
  year={2018}
}

@inproceedings{nguyen2011situating,
  title={Situating the concern for information privacy through an empirical study of responses to video recording},
  author={Nguyen, David H and Bedford, Aurora and Bretana, Alexander Gerard and Hayes, Gillian R},
  booktitle={Proceedings of the SIGCHI Conference on Human Factors in Computing Systems},
  pages={3207--3216},
  year={2011}
}

@inproceedings{gopal2023hidden,
  title={Hidden reality: Caution, your hand gesture inputs in the immersive virtual world are visible to all!},
  author={Gopal, Sindhu Reddy Kalathur and Shukla, Diksha and Wheelock, James David and Saxena, Nitesh},
  booktitle={32nd USENIX security symposium (USENIX Security 23)},
  pages={859--876},
  year={2023}
}

@article{gallardo2023speculative,
  title={Speculative privacy concerns about ar glasses data collection},
  author={Gallardo, Andrea and Choy, Chris and Juneja, Jaideep and Bozkir, Efe and Cobb, Camille and Bauer, Lujo and Cranor, Lorrie},
  journal={Proceedings on Privacy Enhancing Technologies},
  year={2023}
}

@inproceedings{nguyen2009encountering,
  title={Encountering SenseCam: personal recording technologies in everyday life},
  author={Nguyen, David H and Marcu, Gabriela and Hayes, Gillian R and Truong, Khai N and Scott, James and Langheinrich, Marc and Roduner, Christof},
  booktitle={Proceedings of the 11th international conference on Ubiquitous computing},
  pages={165--174},
  year={2009}
}

@inproceedings{procyk2014exploring,
  title={Exploring video streaming in public settings: shared geocaching over distance using mobile video chat},
  author={Procyk, Jason and Neustaedter, Carman and Pang, Carolyn and Tang, Anthony and Judge, Tejinder K},
  booktitle={Proceedings of the SIGCHI Conference on Human Factors in Computing Systems},
  pages={2163--2172},
  year={2014}
}

@inproceedings{krombholz2015ok,
  title={Ok glass, leave me alone: towards a systematization of privacy enhancing technologies for wearable computing},
  author={Krombholz, Katharina and Dabrowski, Adrian and Smith, Matthew and Weippl, Edgar},
  booktitle={International Conference on Financial Cryptography and Data Security},
  pages={274--280},
  year={2015},
  organization={Springer}
}

@inproceedings{koelle2015don,
  title={Don't look at me that way! Understanding user attitudes towards data glasses usage},
  author={Koelle, Marion and Kranz, Matthias and M{\"o}ller, Andreas},
  booktitle={Proceedings of the 17th international conference on human-computer interaction with mobile devices and services},
  pages={362--372},
  year={2015}
}

@inproceedings{portnoff2015somebody,
  title={Somebody's watching me? assessing the effectiveness of webcam indicator lights},
  author={Portnoff, Rebecca S and Lee, Linda N and Egelman, Serge and Mishra, Pratyush and Leung, Derek and Wagner, David},
  booktitle={Proceedings of the 33rd Annual ACM Conference on Human Factors in Computing Systems},
  pages={1649--1658},
  year={2015}
}

@inproceedings{bipat2019analyzing,
  title={Analyzing the use of camera glasses in the wild},
  author={Bipat, Taryn and Bos, Maarten Willem and Vaish, Rajan and Monroy-Hern{\'a}ndez, Andr{\'e}s},
  booktitle={Proceedings of the 2019 CHI Conference on Human Factors in Computing Systems},
  pages={1--8},
  year={2019}
}

@inproceedings{thakkar2022would,
  title={“It would probably turn into a social faux-pas”: Users’ and Bystanders’ Preferences of Privacy Awareness Mechanisms in Smart Homes},
  author={Thakkar, Parth Kirankumar and He, Shijing and Xu, Shiyu and Huang, Danny Yuxing and Yao, Yaxing},
  booktitle={Proceedings of the 2022 CHI Conference on Human Factors in Computing Systems},
  pages={1--13},
  year={2022}
}

@inproceedings{albayaydh2023examining,
  title={Examining power dynamics and user privacy in smart technology use among jordanian households},
  author={Albayaydh, Wael and Flechais, Ivan},
  booktitle={32nd USENIX Security Symposium (USENIX Security 23)},
  pages={4643--4659},
  year={2023}
}

@inproceedings{prange2021priview,
  title={Priview--exploring visualisations to support users’ privacy awareness},
  author={Prange, Sarah and Shams, Ahmed and Piening, Robin and Abdelrahman, Yomna and Alt, Florian},
  booktitle={Proceedings of the 2021 chi conference on human factors in computing systems},
  pages={1--18},
  year={2021}
}

@inproceedings{mcduff2018inphysible,
  title={Inphysible: Camouflage against video-based physiological measurement},
  author={McDuff, Daniel and Hurter, Christophe},
  booktitle={2018 40th Annual International Conference of the IEEE Engineering in Medicine and Biology Society (EMBC)},
  pages={5784--5789},
  year={2018},
  organization={IEEE}
}

@incollection{patel2009blindspot,
  title={Blindspot: Creating capture-resistant spaces},
  author={Patel, Shwetak N and Summet, Jay W and Truong, Khai N},
  booktitle={Protecting Privacy in Video Surveillance},
  pages={185--201},
  year={2009},
  publisher={Springer}
}

@article{ra2017not,
  title={Do not capture: Automated obscurity for pervasive imaging},
  author={Ra, Moo-Ryong and Lee, Seungjoon and Miluzzo, Emiliano and Zavesky, Eric},
  journal={IEEE Internet Computing},
  volume={21},
  number={3},
  pages={82--87},
  year={2017},
  publisher={IEEE}
}

@inproceedings{bo2014privacy,
  title={Privacy. tag: Privacy concern expressed and respected},
  author={Bo, Cheng and Shen, Guobin and Liu, Jie and Li, Xiang-Yang and Zhang, YongGuang and Zhao, Feng},
  booktitle={Proceedings of the 12th ACM conference on embedded network sensor systems},
  pages={163--176},
  year={2014}
}

@inproceedings{marky2020you,
  title={“You just can’t know about everything”: Privacy Perceptions of Smart Home Visitors},
  author={Marky, Karola and Prange, Sarah and Krell, Florian and M{\"u}hlh{\"a}user, Max and Alt, Florian},
  booktitle={Proceedings of the 19th International Conference on Mobile and Ubiquitous Multimedia},
  pages={83--95},
  year={2020}
}

@article{alharbi2019mask,
  title={To mask or not to mask? balancing privacy with visual confirmation utility in activity-oriented wearable cameras},
  author={Alharbi, Rawan and Tolba, Mariam and Petito, Lucia C and Hester, Josiah and Alshurafa, Nabil},
  journal={Proceedings of the ACM on interactive, mobile, wearable and ubiquitous technologies},
  volume={3},
  number={3},
  pages={1--29},
  year={2019},
  publisher={ACM New York, NY, USA}
}

@article{dimiccoli2018mitigating,
  title={Mitigating bystander privacy concerns in egocentric activity recognition with deep learning and intentional image degradation},
  author={Dimiccoli, Mariella and Mar{\'\i}n, Juan and Thomaz, Edison},
  journal={Proceedings of the ACM on Interactive, Mobile, Wearable and Ubiquitous Technologies},
  volume={1},
  number={4},
  pages={1--18},
  year={2018},
  publisher={ACM New York, NY, USA}
}

@inproceedings{roesner2014world,
  title={World-driven access control for continuous sensing},
  author={Roesner, Franziska and Molnar, David and Moshchuk, Alexander and Kohno, Tadayoshi and Wang, Helen J},
  booktitle={Proceedings of the 2014 ACM SIGSAC conference on computer and communications security},
  pages={1169--1181},
  year={2014}
}

@inproceedings{williamson2022digital,
  title={Digital proxemics: Designing social and collaborative interaction in virtual environments},
  author={Williamson, Julie R and O'Hagan, Joseph and Guerra-Gomez, John Alexis and Williamson, John H and Cesar, Pablo and Shamma, David A},
  booktitle={Proceedings of the 2022 CHI conference on human factors in computing systems},
  pages={1--12},
  year={2022}
}

@inproceedings{meirose2018towards,
  title={Towards Respectful Smart Glasses through Conversation Detection.},
  author={Meirose, Franziska and Schultze, Sven and Kuehlewind, Sebastian and Koelle, Marion and Abdenebaoui, Larbi and Boll, Susanne},
  booktitle={MuC},
  year={2018}
}

@inproceedings{wolf2018we,
  title={We should start thinking about privacy implications of sonic input in everyday augmented reality!},
  author={Wolf, Katrin and Marky, Karola and Funk, Markus},
  booktitle={Mensch und Computer 2018-Workshopband},
  pages={10--18420},
  year={2018},
  organization={Gesellschaft f{\"u}r Informatik eV}
}

@inproceedings{steil2019privaceye,
  title={Privaceye: privacy-preserving head-mounted eye tracking using egocentric scene image and eye movement features},
  author={Steil, Julian and Koelle, Marion and Heuten, Wilko and Boll, Susanne and Bulling, Andreas},
  booktitle={Proceedings of the 11th ACM symposium on eye tracking research \& applications},
  pages={1--10},
  year={2019}
}

@article{malhotra2004internet,
  title={Internet users' information privacy concerns (IUIPC): The construct, the scale, and a causal model},
  author={Malhotra, Naresh K and Kim, Sung S and Agarwal, James},
  journal={Information systems research},
  volume={15},
  number={4},
  pages={336--355},
  year={2004},
  publisher={Informs}
}

@article{rogers1975protection,
  title={A protection motivation theory of fear appeals and attitude change1},
  author={Rogers, Ronald W},
  journal={The journal of psychology},
  volume={91},
  number={1},
  pages={93--114},
  year={1975},
  publisher={Taylor \& Francis}
}

@article{maddux1983protection,
  title={Protection motivation and self-efficacy: A revised theory of fear appeals and attitude change},
  author={Maddux, James E and Rogers, Ronald W},
  journal={Journal of experimental social psychology},
  volume={19},
  number={5},
  pages={469--479},
  year={1983},
  publisher={Elsevier}
}

@article{nissenbaum2004privacy,
  title={Privacy as contextual integrity},
  author={Nissenbaum, Helen},
  journal={Wash. L. Rev.},
  volume={79},
  pages={119},
  year={2004},
  publisher={HeinOnline}
}

@inproceedings{windl2023understanding,
  title={Understanding and mitigating technology-facilitated privacy violations in the physical world},
  author={Windl, Maximiliane and Winterhalter, Verena and Schmidt, Albrecht and Mayer, Sven},
  booktitle={Proceedings of the 2023 CHI Conference on Human Factors in Computing Systems},
  pages={1--16},
  year={2023}
}

@inproceedings{chung2023negotiating,
  title={Negotiating dyadic interactions through the lens of augmented reality glasses},
  author={Chung, Ji Won and Fu, Xiyu Jenny and Deocadiz-Smith, Zachary and Jung, Malte F and Huang, Jeff},
  booktitle={Proceedings of the 2023 ACM Designing Interactive Systems Conference},
  pages={493--508},
  year={2023}
}

@inproceedings{hoyle2014privacy,
  title={Privacy behaviors of lifeloggers using wearable cameras},
  author={Hoyle, Roberto and Templeman, Robert and Armes, Steven and Anthony, Denise and Crandall, David and Kapadia, Apu},
  booktitle={Proceedings of the 2014 ACM international joint conference on pervasive and ubiquitous computing},
  pages={571--582},
  year={2014}
}

@misc{xromXiaomiRaises,
	author = {VRAR World},
	title = {{X}iaomi {R}aises {A}{I} {S}mart {G}lasses {S}ales {T}arget to 500,000 {U}nits.},
	howpublished = {\url{https://www.xrom.in/post/xiaomi-raises-ai-smart-glasses-sales-target-to-500-000-units}},
	year = {2025},
	note = {[Accessed 19-08-2025]},
}

@misc{xromForecasts107,
	title = {IDC Forecasts 107\% Year-on-Year Growth for Chinas Smart Glasses Shipments, Projected to Reach 2.75 Million Units in 2025},
        author = {World, VRAR},
	howpublished = {\url{https://www.xrom.in/post/idc-forecasts-107-year-on-year-growth-for-chinas-smart-glasses-shipments-projected-to-reach-2-75-mi}},
	year = {2025},
	note = {[Accessed 19-08-2025]},
}

@inproceedings{alshehri2023exploring,
  title={Exploring the negotiation behaviors of owners and bystanders over data practices of smart home devices},
  author={Alshehri, Ahmed and Pahk, Eugin and Spielman, Joseph and Parker, Jacob T and Gilbert, Benjamin and Yue, Chuan},
  booktitle={Proceedings of the 2023 CHI Conference on Human Factors in Computing Systems},
  pages={1--27},
  year={2023}
}

@inproceedings{wobbrock2011aligned,
  title={The aligned rank transform for nonparametric factorial analyses using only anova procedures},
  author={Wobbrock, Jacob O and Findlater, Leah and Gergle, Darren and Higgins, James J},
  booktitle={Proceedings of the SIGCHI conference on human factors in computing systems},
  pages={143--146},
  year={2011}
}

@inproceedings{elkin2021aligned,
  title={An aligned rank transform procedure for multifactor contrast tests},
  author={Elkin, Lisa A and Kay, Matthew and Higgins, James J and Wobbrock, Jacob O},
  booktitle={The 34th annual ACM symposium on user interface software and technology},
  pages={754--768},
  year={2021}
}

@article{kokolakis2017privacy,
  title={Privacy attitudes and privacy behaviour: A review of current research on the privacy paradox phenomenon},
  author={Kokolakis, Spyros},
  journal={Computers \& security},
  volume={64},
  pages={122--134},
  year={2017},
  publisher={Elsevier}
}

@article{page2021prisma,
  title={The PRISMA 2020 statement: an updated guideline for reporting systematic reviews},
  author={Page, Matthew J and McKenzie, Joanne E and Bossuyt, Patrick M and Boutron, Isabelle and Hoffmann, Tammy C and Mulrow, Cynthia D and Shamseer, Larissa and Tetzlaff, Jennifer M and Akl, Elie A and Brennan, Sue E and others},
  journal={bmj},
  volume={372},
  year={2021},
  publisher={British Medical Journal Publishing Group}
}

@article{thomas2008methods,
  title={Methods for the thematic synthesis of qualitative research in systematic reviews},
  author={Thomas, James and Harden, Angela},
  journal={BMC medical research methodology},
  volume={8},
  number={1},
  pages={45},
  year={2008},
  publisher={Springer}
}

@misc{spectaclesMessages,
	author = {Snap Inc.},
	title = {{L}{E}{D} {M}essages --- support.spectacles.com},
	howpublished = {\url{https://support.spectacles.com/hc/en-us/articles/360033763171-LED-Messages}},
	year = {2025},
	note = {[Accessed 04-09-2025]},
}

@misc{wiredKoreaBeeping,
	author = {Reuters},
	title = {{K}orea: {B}eeping {P}revents {P}eeping --- wired.com},
	howpublished = {\url{https://www.wired.com/2003/11/korea-beeping-prevents-peeping}},
	year = {2003},
	note = {[Accessed 04-09-2025]},
}

@misc{eyesightAVP,
	author = {Apple Inc.},
	title = {What EyeSight shows on Apple Vision Pro},
	howpublished = {\url{https://support.apple.com/en-us/120481}},
	year = {2024},
	note = {[Accessed 04-09-2025]},
}

@inproceedings{teyssier2021eyecam,
  title={Eyecam: Revealing relations between humans and sensing devices through an anthropomorphic webcam},
  author={Teyssier, Marc and Koelle, Marion and Strohmeier, Paul and Fruchard, Bruno and Steimle, J{\"u}rgen},
  booktitle={Proceedings of the 2021 CHI Conference on Human Factors in Computing Systems},
  pages={1--13},
  year={2021}
}

@inproceedings{koelle2019evaluating,
  title={Evaluating a Wearable Camera's Social Acceptability In-the-Wild},
  author={Koelle, Marion and Wallbaum, Torben and Heuten, Wilko and Boll, Susanne},
  booktitle={Extended Abstracts of the 2019 CHI Conference on Human Factors in Computing Systems},
  pages={1--6},
  year={2019}
}

@inproceedings{escher2022transparency,
  title={Transparency for Bystanders in IoT regarding audiovisual Recordings},
  author={Escher, Stephan and Etzrodt, Katrin and Weller, Benjamin and K{\"o}psell, Stefan and Strufe, Thorsten},
  booktitle={2022 IEEE International Conference on Pervasive Computing and Communications Workshops and other Affiliated Events (PerCom Workshops)},
  pages={649--654},
  year={2022},
  organization={IEEE}
}

@inproceedings{barhm2011negotiating,
  title={Negotiating privacy preferences in video surveillance systems},
  author={Barhm, Mukhtaj S and Qwasmi, Nidal and Qureshi, Faisal Z and El-Khatib, Khalil},
  booktitle={International Conference on Industrial, Engineering and Other Applications of Applied Intelligent Systems},
  pages={511--521},
  year={2011},
  organization={Springer}
}

@inproceedings{ashok2014not,
  title={Do not share! Invisible light beacons for signaling preferences to privacy-respecting cameras},
  author={Ashok, Ashwin and Nguyen, Viet and Gruteser, Marco and Mandayam, Narayan and Yuan, Wenjia and Dana, Kristin},
  booktitle={Proceedings of the 1st ACM MobiCom workshop on Visible light communication systems},
  pages={39--44},
  year={2014}
}

@inproceedings{liao2025bystander,
  title={Bystander Privacy in Video Sharing Era: Automated Consent Compliance through Platform Censorship},
  author={Liao, Si and He, Hanwei and Chen, Huangxun and Yang, Zhice},
  booktitle={Proceedings of the 2025 CHI Conference on Human Factors in Computing Systems},
  pages={1--16},
  year={2025}
}

@inproceedings{kim2023erebus,
  title={Erebus: Access control for augmented reality systems},
  author={Kim, Yoonsang and Goutam, Sanket and Rahmati, Amir and Kaufman, Arie},
  booktitle={32nd USENIX Security Symposium (USENIX Security 23)},
  pages={929--946},
  year={2023}
}

@inproceedings{zhou2024bring,
  title={Bring privacy to the table: Interactive negotiation for privacy settings of shared sensing devices},
  author={Zhou, Haozhe and Goel, Mayank and Agarwal, Yuvraj},
  booktitle={Proceedings of the 2024 CHI Conference on Human Factors in Computing Systems},
  pages={1--22},
  year={2024}
}

@inproceedings{templeman2014placeavoider,
  title={PlaceAvoider: Steering First-Person Cameras away from Sensitive Spaces.},
  author={Templeman, Robert and Korayem, Mohammed and Crandall, David J and Kapadia, Apu},
  booktitle={NDSS},
  volume={14},
  pages={23--26},
  year={2014}
}

@inproceedings{henne2013snapme,
  title={SnapMe if you can: Privacy threats of other peoples' geo-tagged media and what we can do about it},
  author={Henne, Benjamin and Szongott, Christian and Smith, Matthew},
  booktitle={Proceedings of the sixth ACM conference on Security and privacy in wireless and mobile networks},
  pages={95--106},
  year={2013}
}

@article{zhang2018cloak,
  title={Cloak of invisibility: Privacy-friendly photo capturing and sharing system},
  author={Zhang, Lan and Li, Xiang-Yang and Liu, Kebin and Liu, Cihang and Ding, Xuan and Liu, Yunhao},
  journal={IEEE Transactions on Mobile Computing},
  volume={18},
  number={11},
  pages={2488--2501},
  year={2018},
  publisher={IEEE}
}

@inproceedings{li2019hideme,
  title={Hideme: Privacy-preserving photo sharing on social networks},
  author={Li, Fenghua and Sun, Zhe and Li, Ang and Niu, Ben and Li, Hui and Cao, Guohong},
  booktitle={IEEE INFOCOM 2019-IEEE Conference on Computer Communications},
  pages={154--162},
  year={2019},
  organization={IEEE}
}

@article{mccann2003grounded,
  title={Grounded theory in nursing research: Part 1--Methodology},
  author={McCann, Terence V and Clark, Eileen},
  year={2003},
  publisher={Nurse Researcher}
}

@book{blandford2016qualitative,
  title={Qualitative HCI research: Going behind the scenes},
  author={Blandford, Ann and Furniss, Dominic and Makri, Stephann},
  year={2016},
  publisher={Morgan \& Claypool Publishers}
}

@article{mayring2014qualitative,
  title={Qualitative content analysis: theoretical foundation, basic procedures and software solution},
  author={Mayring, Philipp},
  year={2014},
  publisher={AUT}
}

@inproceedings{abraham2024don,
  title={Don’t Record My Private pARts: Understanding The Role of Sensitive Contexts and Privacy Perceptions in Influencing Attitudes Towards Everyday Augmented Reality Sensor Usage},
  author={Abraham, Melvin and Khamis, Mohamed and McGill, Mark},
  booktitle={2024 IEEE International Symposium on Mixed and Augmented Reality (ISMAR)},
  pages={749--758},
  year={2024},
  organization={IEEE}
}

@article{cruz2025assessing,
  title={Assessing User Perceptions and Preferences on Applying Obfuscation Techniques for Privacy Protection in Augmented Reality},
  author={Cruz, Ana Cassia and Costa, Rog{\'e}rio Lu{\'\i}s de C and Santos, Leonel and Rabad{\~a}o, Carlos and Marto, Anabela and Gon{\c{c}}alves, Alexandrino},
  journal={Future Internet},
  volume={17},
  number={2},
  pages={55},
  year={2025},
  publisher={MDPI}
}

@inproceedings{mujirishvili2024don,
  title={“I Don’t Want to Become a Number’’: Examining Different Stakeholder Perspectives on a Video-Based Monitoring System for Senior Care with Inherent Privacy Protection (by Design).},
  author={Mujirishvili, Tamara and Fedosov, Anton and Hashemifard, Kooshan and Climent-P{\'e}rez, Pau and Florez-Revuelta, Francisco},
  booktitle={Proceedings of the 2024 CHI Conference on Human Factors in Computing Systems},
  pages={1--19},
  year={2024}
}

@article{do2021smart,
  title={Smart webcam cover: Exploring the design of an intelligent webcam cover to improve usability and trust},
  author={Do, Youngwook and Park, Jung Wook and Wu, Yuxi and Basu, Avinandan and Zhang, Dingtian and Abowd, Gregory D and Das, Sauvik},
  journal={Proceedings of the ACM on Interactive, Mobile, Wearable and Ubiquitous Technologies},
  volume={5},
  number={4},
  pages={1--21},
  year={2021},
  publisher={ACM New York, NY, USA}
}

@inproceedings{yang2024feasibility,
  title={On the Feasibility of Predicting Users' Privacy Concerns using Contextual Labels and Personal Preferences},
  author={Yang, Yaqing and Li, Tony W and Jin, Haojian},
  booktitle={Proceedings of the 2024 CHI Conference on Human Factors in Computing Systems},
  pages={1--20},
  year={2024}
}

@inproceedings{rajaram2025exploring,
  title={Exploring the Design Space of Privacy-Driven Adaptation Techniques for Future Augmented Reality Interfaces},
  author={Rajaram, Shwetha and Peralta, Macarena and Johnson, Janet G and Nebeling, Michael},
  booktitle={Proceedings of the 2025 CHI Conference on Human Factors in Computing Systems},
  pages={1--19},
  year={2025}
}

@inproceedings{windl2025designing,
  title={Designing Effective Consent Mechanisms for Spontaneous Interactions in Augmented Reality},
  author={Windl, Maximiliane and Laboda, Petra Zsofia and Mayer, Sven},
  booktitle={Proceedings of the 2025 CHI Conference on Human Factors in Computing Systems},
  pages={1--18},
  year={2025}
}

@article{venugopalan2024aragorn,
  title={Aragorn: A privacy-enhancing system for mobile cameras},
  author={Venugopalan, Hari and Din, Zainul Abi and Carpenter, Trevor and Lowe-Power, Jason and King, Samuel T and Shafiq, Zubair},
  journal={Proceedings of the ACM on Interactive, Mobile, Wearable and Ubiquitous Technologies},
  volume={7},
  number={4},
  pages={1--31},
  year={2024},
  publisher={ACM New York, NY, USA}
}

@article{fisher1993social,
  title={Social desirability bias and the validity of indirect questioning},
  author={Fisher, Robert J},
  journal={Journal of consumer research},
  volume={20},
  number={2},
  pages={303--315},
  year={1993},
  publisher={The University of Chicago Press}
}

@article{miltgen2014cultural,
  title={Cultural and generational influences on privacy concerns: a qualitative study in seven European countries},
  author={Miltgen, Caroline Lancelot and Peyrat-Guillard, Dominique},
  journal={European journal of information systems},
  volume={23},
  number={2},
  pages={103--125},
  year={2014},
  publisher={Taylor \& Francis}
}

@article{joinson2010privacy,
  title={Privacy, trust, and self-disclosure online},
  author={Joinson, Adam N and Reips, Ulf-Dietrich and Buchanan, Tom and Schofield, Carina B Paine},
  journal={Human--Computer Interaction},
  volume={25},
  number={1},
  pages={1--24},
  year={2010},
  publisher={Taylor \& Francis}
}

@inproceedings{olson2005study,
  title={A study of preferences for sharing and privacy},
  author={Olson, Judith S and Grudin, Jonathan and Horvitz, Eric},
  booktitle={CHI'05 extended abstracts on Human factors in computing systems},
  pages={1985--1988},
  year={2005}
}

@article{altman1975environment,
  title={The environment and social behavior: privacy, personal space, territory, and crowding.},
  author={Altman, Irwin},
  year={1975},
  publisher={ERIC}
}

@article{altman1977privacy,
  title={Privacy regulation: Culturally universal or culturally specific?},
  author={Altman, Irwin},
  journal={Journal of social issues},
  volume={33},
  number={3},
  pages={66--84},
  year={1977},
  publisher={Wiley Online Library}
}

@article{rao2020types,
  title={Types of privacy expectations},
  author={Rao, Ashwini and Pfeffer, Juergen},
  journal={Frontiers in big Data},
  volume={3},
  pages={7},
  year={2020},
  publisher={Frontiers Media SA}
}

@article{zhang2025can,
  title={Can you see what I see? Examining the Impact of Smart Glasses on Communication Dynamics in Distributed Emergency Medical Teams},
  author={Zhang, Zhan and Bai, Enze and Xu, Yincao and Adelgais, Kathleen and Ozkaynak, Mustafa},
  journal={Proceedings of the ACM on Human-Computer Interaction},
  volume={9},
  number={7},
  pages={1--34},
  year={2025},
  publisher={ACM New York, NY, USA}
}

@article{nicholas2022friendscope,
  title={Friendscope: Exploring In-the-Moment Experience Sharing on Camera Glasses via a Shared Camera},
  author={Nicholas, Molly Jane and Smith, Brian A and Vaish, Rajan},
  journal={Proceedings of the ACM on Human-Computer Interaction},
  volume={6},
  number={CSCW1},
  pages={1--25},
  year={2022},
  publisher={ACM New York, NY, USA}
}

@inproceedings{huh2025vid2coach,
  title={Vid2Coach: Transforming How-To Videos into Task Assistants},
  author={Huh, Mina and Xue, Zihui and Das, Ujjaini and Ashutosh, Kumar and Grauman, Kristen and Pavel, Amy},
  booktitle={Proceedings of the 38th Annual ACM Symposium on User Interface Software and Technology},
  pages={1--24},
  year={2025}
}

@article{chang2020medglasses,
  title={MedGlasses: A wearable smart-glasses-based drug pill recognition system using deep learning for visually impaired chronic patients},
  author={Chang, Wan-Jung and Chen, Liang-Bi and Hsu, Chia-Hao and Chen, Jheng-Hao and Yang, Tzu-Chin and Lin, Cheng-Pei},
  journal={IEEE Access},
  volume={8},
  pages={17013--17024},
  year={2020},
  publisher={IEEE}
}

@inproceedings{windl2022automating,
  title={Automating contextual privacy policies: Design and evaluation of a production tool for digital consumer privacy awareness},
  author={Windl, Maximiliane and Henze, Niels and Schmidt, Albrecht and Feger, Sebastian S},
  booktitle={Proceedings of the 2022 CHI Conference on Human Factors in Computing Systems},
  pages={1--18},
  year={2022}
}

@inproceedings{tran2025wearable,
  title={Wearable AR in Everyday Contexts: Insights from a Digital Ethnography of YouTube Videos},
  author={Tran, Tram Thi Minh and Brown, Shane and Weidlich, Oliver and Yoo, Soojeong and Parker, Callum},
  booktitle={Proceedings of the 2025 CHI Conference on Human Factors in Computing Systems},
  pages={1--18},
  year={2025}
}

@article{iqbal2023adopting,
  title={Adopting smart glasses responsibly: potential benefits, ethical, and privacy concerns with Ray-Ban stories},
  author={Iqbal, Muhammad Zahid and Campbell, Abraham G},
  journal={AI and Ethics},
  volume={3},
  number={1},
  pages={325--327},
  year={2023},
  publisher={Springer}
}

@inproceedings{rajaram2025privacy,
  title={Privacy Equilibrium: Balancing Privacy Needs in Dynamic Multi-User Augmented Reality Scenarios},
  author={Rajaram, Shwetha and Chen, Jiasi and Nebeling, Michael},
  booktitle={Proceedings of the 38th Annual ACM Symposium on User Interface Software and Technology},
  pages={1--24},
  year={2025}
}

@inproceedings{jimenez2014tag,
  title={Tag detection for preventing unauthorized face image processing},
  author={Jimenez, Alberto Escalada and Dabrowski, Adrian and Sonehara, Noburu and Martinez, Juan M Montero and Echizen, Isao},
  booktitle={International Workshop on Digital Watermarking},
  pages={513--524},
  year={2014},
  organization={Springer}
}

@article{eisenberg1988order,
  title={Order effects: A study of the possible influence of presentation order on user judgments of document relevance},
  author={Eisenberg, Michael and Barry, Carol},
  journal={Journal of the American Society for Information Science},
  volume={39},
  number={5},
  pages={293--300},
  year={1988},
  publisher={Wiley Online Library}
}

@book{lazar2017research,
  title={Research methods in human-computer interaction},
  author={Lazar, Jonathan and Feng, Jinjuan Heidi and Hochheiser, Harry},
  year={2017},
  publisher={Morgan Kaufmann}
}

@article{kudina2019ethics,
  title={Ethics from within: Google Glass, the Collingridge dilemma, and the mediated value of privacy},
  author={Kudina, Olya and Verbeek, Peter-Paul},
  journal={Science, Technology, \& Human Values},
  volume={44},
  number={2},
  pages={291--314},
  year={2019},
  publisher={SAGE Publications Sage CA: Los Angeles, CA}
}

@inproceedings{wiese2011you,
  title={Are you close with me? Are you nearby? Investigating social groups, closeness, and willingness to share},
  author={Wiese, Jason and Kelley, Patrick Gage and Cranor, Lorrie Faith and Dabbish, Laura and Hong, Jason I and Zimmerman, John},
  booktitle={Proceedings of the 13th international conference on Ubiquitous computing},
  pages={197--206},
  year={2011}
}

@article{fogel2009internet,
  title={Internet social network communities: Risk taking, trust, and privacy concerns},
  author={Fogel, Joshua and Nehmad, Elham},
  journal={Computers in human behavior},
  volume={25},
  number={1},
  pages={153--160},
  year={2009},
  publisher={Elsevier}
}

@inproceedings{stutzman2010friends,
  title={Friends only: examining a privacy-enhancing behavior in facebook},
  author={Stutzman, Fred and Kramer-Duffield, Jacob},
  booktitle={Proceedings of the SIGCHI conference on human factors in computing systems},
  pages={1553--1562},
  year={2010}
}

@inproceedings{akter2020uncomfortable,
  title={" I am uncomfortable sharing what I can't see": Privacy Concerns of the Visually Impaired with Camera Based Assistive Applications},
  author={Akter, Taslima and Dosono, Bryan and Ahmed, Tousif and Kapadia, Apu and Semaan, Bryan},
  booktitle={29th USENIX Security Symposium (USENIX Security 20)},
  pages={1929--1948},
  year={2020}
}

@inproceedings{zhao2023if,
  title={$\{$“If$\}$ sighted people know, I should be able to $\{$know:”$\}$ Privacy Perceptions of Bystanders with Visual Impairments around Camera-based Technology},
  author={Zhao, Yuhang and Yao, Yaxing and Fu, Jiaru and Zhou, Nihan},
  booktitle={32nd USENIX Security Symposium (USENIX Security 23)},
  pages={4661--4678},
  year={2023}
}

@inproceedings{singhal2016you,
  title={You are being watched: Bystanders' perspective on the use of camera devices in public spaces},
  author={Singhal, Samarth and Neustaedter, Carman and Schiphorst, Thecla and Tang, Anthony and Patra, Abhisekh and Pan, Rui},
  booktitle={Proceedings of the 2016 CHI Conference Extended Abstracts on Human Factors in Computing Systems},
  pages={3197--3203},
  year={2016}
}

@inproceedings{rajaram2023reframe,
  title={Reframe: An augmented reality storyboarding tool for character-driven analysis of security \& privacy concerns},
  author={Rajaram, Shwetha and Roesner, Franziska and Nebeling, Michael},
  booktitle={Proceedings of the 36th annual ACM symposium on user interface software and technology},
  pages={1--15},
  year={2023}
}

@inproceedings{al2025bystandaria,
  title={BystandARIA: Enabling AR Bystander Privacy using LEDs},
  author={Al Aaraj, Jad and Markopoulou, Athina},
  booktitle={Proceedings of the Twenty-sixth International Symposium on Theory, Algorithmic Foundations, and Protocol Design for Mobile Networks and Mobile Computing},
  pages={444--449},
  year={2025}
}

@article{ghaiumy2021difficulties,
  title={Difficulties of measuring culture in privacy studies},
  author={Ghaiumy Anaraky, Reza and Li, Yao and Knijnenburg, Bart},
  journal={Proceedings of the ACM on Human-Computer Interaction},
  volume={5},
  number={CSCW2},
  pages={1--26},
  year={2021},
  publisher={ACM New York, NY, USA}
}

@incollection{roberts2017privacy,
  title={Privacy: A cultural view},
  author={Roberts, John M and Gregor, Thomas},
  booktitle={Privacy and Personality},
  pages={199--225},
  year={2017},
  publisher={Routledge}
}

@article{kim2025cultural,
  title={Cultural Differences in the Use of Augmented Reality Smart Glasses (ARSGs) Between the US and South Korea: Privacy Concerns and the Technology Acceptance Model},
  author={Kim, Se Jung and Lee, Yoon Esther and Chock, T Makana},
  journal={Applied Sciences},
  volume={15},
  number={13},
  pages={7430},
  year={2025},
  publisher={MDPI}
}

@article{li2022cultural,
  title={Cultural differences in the effects of contextual factors and privacy concerns on users’ privacy decision on social networking sites},
  author={Li, Yao and Rho, Eugenia Ha Rim and Kobsa, Alfred},
  journal={Behaviour \& Information Technology},
  volume={41},
  number={3},
  pages={655--677},
  year={2022},
  publisher={Taylor \& Francis}
}

@inproceedings{marky2020don,
  title={” I don’t know how to protect myself”: Understanding Privacy Perceptions Resulting from the Presence of Bystanders in Smart Environments},
  author={Marky, Karola and Voit, Alexandra and St{\"o}ver, Alina and Kunze, Kai and Schr{\"o}der, Svenja and M{\"u}hlh{\"a}user, Max},
  booktitle={Proceedings of the 11th Nordic Conference on Human-Computer Interaction: Shaping Experiences, Shaping Society},
  pages={1--11},
  year={2020}
}

@article{ahmed2018up,
  title={Up to a limit? privacy concerns of bystanders and their willingness to share additional information with visually impaired users of assistive technologies},
  author={Ahmed, Tousif and Kapadia, Apu and Potluri, Venkatesh and Swaminathan, Manohar},
  journal={Proceedings of the ACM on Interactive, Mobile, Wearable and Ubiquitous Technologies},
  volume={2},
  number={3},
  pages={1--27},
  year={2018},
  publisher={ACM New York, NY, USA}
}

@article{draper2017privacy,
  title={From privacy pragmatist to privacy resigned: Challenging narratives of rational choice in digital privacy debates},
  author={Draper, Nora A},
  journal={Policy \& Internet},
  volume={9},
  number={2},
  pages={232--251},
  year={2017},
  publisher={Wiley Online Library}
}

@article{egelman2015predicting,
  title={Predicting privacy and security attitudes},
  author={Egelman, Serge and Peer, Eyal},
  journal={ACM SIGCAS computers and society},
  volume={45},
  number={1},
  pages={22--28},
  year={2015},
  publisher={ACM New York, NY, USA}
}

@inproceedings{egelman2016behavior,
  title={Behavior ever follows intention? A validation of the Security Behavior Intentions Scale (SeBIS)},
  author={Egelman, Serge and Harbach, Marian and Peer, Eyal},
  booktitle={Proceedings of the 2016 CHI conference on human factors in computing systems},
  pages={5257--5261},
  year={2016}
}

@inproceedings{shu2018cardea,
  title={Cardea: Context-aware visual privacy protection for photo taking and sharing},
  author={Shu, Jiayu and Zheng, Rui and Hui, Pan},
  booktitle={Proceedings of the 9th ACM Multimedia Systems Conference},
  pages={304--315},
  year={2018}
}

@inproceedings{al2021role,
  title={The role of citizens’ familiarity, privacy concerns, and trust on adoption of smart services},
  author={Al-Musawi, Ali and Yang, Erik and Bley, Katja and Thapa, Devinder and Pappas, Ilias O},
  booktitle={Norsk IKT-konferanse for forskning og utdanning},
  number={2},
  year={2021}
}

@article{baruh2014more,
  title={It is more than personal: Development and validation of a multidimensional privacy orientation scale},
  author={Baruh, Lemi and Cemalc{\i}lar, Zeynep},
  journal={Personality and Individual Differences},
  volume={70},
  pages={165--170},
  year={2014},
  publisher={Elsevier}
}

@inproceedings{hasan2023psychometric,
  title={A psychometric scale to measure individuals’ value of other people’s privacy (VOPP)},
  author={Hasan, Rakibul and Weil, Rebecca and Siegel, Rudolf and Krombholz, Katharina},
  booktitle={Proceedings of the 2023 chi conference on human factors in computing systems},
  pages={1--14},
  year={2023}
}

@article{lee2021structural,
  title={Structural model of the healthcare information security behavior of nurses applying protection motivation theory},
  author={Lee, EunWon and Seomun, GyeongAe},
  journal={International journal of environmental research and public health},
  volume={18},
  number={4},
  pages={2084},
  year={2021},
  publisher={MDPI}
}

@article{ma2015survey,
  title={A survey-based study of factors that motivate nurses to protect the privacy of electronic medical records},
  author={Ma, Chen-Chung and Kuo, Kuang-Ming and Alexander, Judith W},
  journal={BMC medical informatics and decision making},
  volume={16},
  number={1},
  pages={13},
  year={2015},
  publisher={Springer}
}

@article{carroll2002adoption,
  title={'No'to a free mobile: when adoption is not enough},
  author={Carroll, Jennie and Howard, Steve and Murphy, Jane and Peck, John},
  year={2002}
}

\appendix

\section{Study 1 Survey Instruments}
\label{appendix:survey}

This appendix presents the complete survey instruments used in Study 1. Both surveys were administered in Chinese and translated to English for presentation. Items were rated on 7-point Likert scales unless otherwise specified.

\subsection{Baseline Attitude Scales}

\subsubsection{Bystander Privacy Concerns Scale}
\label{appendix:bystander-baseline}

Adapted from the Internet Users' Information Privacy Concerns (IUIPC) scale \cite{malhotra2004internet} to assess bystanders' baseline privacy attitudes toward smart glasses recording. Participants rated their agreement from 1 (Strongly Disagree) to 7 (Strongly Agree).

\begin{table*}[h]
\centering
\caption{Bystander Privacy Concerns Scale Items}
\label{tab:bystander-scale}
\begin{tabular}{p{2.5cm}p{10cm}}
\toprule
\textbf{Dimension} & \textbf{Item} \\
\midrule
\multirow{2}{*}{Awareness} 
& AW1: I pay attention to whether I am being photographed or recognized by smart glasses without being informed. \\
& AW2: I pay attention to whether I am being photographed or recognized by smart glasses without my consent. \\
\midrule
\multirow{2}{*}{Control} 
& CT1: I should have the right to decide whether to be photographed by smart glasses and how my image is used. \\
& CT2: If I notice I might be photographed by smart glasses, I would take action to avoid entering the frame or decline recording. \\
\midrule
\multirow{2}{*}{Collection} 
& CL1: When I notice I might be photographed or recognized by smart glasses, I pay attention to potential implications for my personal privacy. \\
& CL2: I pay attention to whether my image captured by smart glasses might be used or shared by others. \\
\bottomrule
\end{tabular}
\end{table*}

\subsubsection{Wearer Privacy Responsibility Scale}
\label{appendix:wearer-baseline}

Developed based on Protection Motivation Theory \cite{rogers1975protection, maddux1983protection} to assess wearers' baseline attitudes toward protecting bystanders' privacy. Participants rated their agreement from 1 (Strongly Disagree) to 7 (Strongly Agree).

\begin{table*}[h]
\centering
\caption{Wearer Privacy Responsibility Scale Items}
\label{tab:wearer-scale}
\begin{tabular}{p{2.5cm}p{10cm}}
\toprule
\textbf{Dimension} & \textbf{Item} \\
\midrule
\multirow{2}{*}{\parbox{2.5cm}{Perceived\\Responsibility}} 
& PR1: I have a responsibility to protect the privacy of people around me when using smart glasses. \\
& PR2: When using smart glasses in public places, I consider whether my recording behavior might cause dissatisfaction or other consequences for others. \\
\midrule
\multirow{2}{*}{\parbox{2.5cm}{Information\\Sharing Intention}} 
& IS1: When using smart glasses' recording or recognition functions, I prefer to inform those around me. \\
& IS2: If people around me want to understand the recording situation, I am willing to inform them of my recording purpose and data usage. \\
\midrule
\multirow{2}{*}{\parbox{2.5cm}{Privacy\\Protection\\Intention}} 
& PP1: Before using smart glasses for recording or recognition, I should obtain consent from relevant people. \\
& PP2: If people around me express concerns or objections, I am willing to stop or adjust my recording behavior. \\
\bottomrule
\end{tabular}
\end{table*}

\subsection{Contextual Scenario Descriptions}
\label{appendix:scenarios}

Participants evaluated six scenarios systematically varying across physical setting (public, semi-public, private/sensitive) and social relationship (acquaintance, stranger). Scenarios were presented in randomized order. For bystanders, scenarios described encountering others using smart glasses; for wearers, scenarios described using smart glasses themselves.

\begin{table*}[h]
\centering
\caption{Contextual Scenario Design}
\label{tab:scenarios}
\begin{tabular}{llll}
\toprule
\textbf{Scenario} & \textbf{Setting} & \textbf{Relationship} & \textbf{Context Description} \\
\midrule
Street & Public & Acquaintance & Recording travel scenery with companions \\
Mall & Public & Stranger & Recording shopping vlog in crowded space \\
Meeting & Semi-public & Acquaintance & Recording meeting for personal minutes \\
Hospital & Semi-public & Stranger & Recording navigation and procedures \\
Private Party & Private & Acquaintance & Recording social gathering at home \\
Gym & Sensitive & Stranger & Recording workout with others present \\
\bottomrule
\end{tabular}
\end{table*}

\subsection{Contextual Measurement Items}
\label{appendix:contextual-items}

For each scenario, participants responded to the following items. Bystanders rated their expectations/needs; wearers rated their willingness to provide.

\subsubsection{Privacy Concern and Recording Reasonability}

\begin{itemize}
    \item \textbf{Bystanders:} ``In this scenario, to what extent would you be concerned about your privacy?'' (1 = Not Concerned at All, 7 = Highly Concerned)
    \item \textbf{Wearers:} ``In this scenario, how reasonable do you think it is to use smart glasses for recording?'' (1 = Completely Unreasonable, 7 = Highly Reasonable)
    \item \textbf{Wearers:} ``In this scenario, to what extent would you be concerned about affecting the privacy of those being recorded?'' (1 = Not Concerned at All, 7 = Highly Concerned)
\end{itemize}

\subsubsection{Information Transparency Dimensions}

Bystanders rated their need for information (1 = Do Not Need at All, 7 = Strongly Need); wearers rated their willingness to disclose (1 = Very Unwilling, 7 = Very Willing).

\begin{table*}[h]
\centering
\caption{Information Transparency Items}
\label{tab:info-items}
\begin{tabular}{ll}
\toprule
\textbf{Dimension} & \textbf{Item} \\
\midrule
Purpose & Purpose and intended use of recording \\
Sharing & Whether data will be uploaded, shared, or made public \\
AI Use & Whether data will be analyzed by AI or algorithms \\
Retention & Data storage method and retention duration \\
Content & Specific content of the recording \\
\bottomrule
\end{tabular}
\end{table*}

\subsubsection{Protective Measure Dimensions}

Bystanders rated their expectations (1 = Do Not Need at All, 7 = Strongly Need); wearers rated their willingness to adopt (1 = Very Unwilling, 7 = Very Willing).

\begin{table*}[h]
\centering
\caption{Protective Measure Items}
\label{tab:protection-items}
\begin{tabular}{ll}
\toprule
\textbf{Measure} & \textbf{Item} \\
\midrule
Proactive Notification & Proactively notify about recording behavior \\
Privacy Filter & Apply privacy protection (e.g., automatic face blurring) \\
No Sharing & Ensure recorded data will not be uploaded or shared \\
Auto Delete & Ensure recorded data will be automatically deleted after a period \\
Prior Consent & Only record after obtaining consent \\
\bottomrule
\end{tabular}
\end{table*}

\subsubsection{Behavioral Response (Bystanders Only)}

Bystanders selected anticipated responses if they discovered being recorded (multiple selections allowed):
\begin{itemize}
    \item Take no action
    \item Try to avoid the camera
    \item Use gestures or actions to express discomfort
    \item Directly ask the person to stop recording
    \item Request deletion of data containing them
    \item File a complaint or report to authorities
\end{itemize}

\subsection{LED Indicator Evaluation}
\label{appendix:led-eval}

Both groups evaluated current LED notification mechanisms.

\textbf{Adequacy Assessment:} ``Most current smart glasses use LED indicators to signal recording. Do you think this method adequately protects [bystanders' privacy / the privacy of those around you]?'' (1 = Very Insufficient, 5 = Very Sufficient)

\textbf{Reasons for Insufficiency} (multiple selections allowed):
\begin{itemize}
    \item Light too small and easily overlooked by people around
    \item Invisible in bright environments
    \item Bystander unfamiliar with LED meaning cannot recognize it
    \item Can be blocked or modified by users
    \item Cannot distinguish between photo, video, or livestream modes
    \item Recording range and resolution unclear
    \item Other: \underline{\hspace{3cm}}
\end{itemize}

\textbf{Preferred Notification Methods} (multiple selections allowed):
\begin{itemize}
    \item No additional methods needed
    \item Audio notifications (shutter sound or voice prompt)
    \item Push notifications to nearby smartphones
    \item Verbal notification from recorder
    \item More visible visual indicators
    \item Other: \underline{\hspace{3cm}}
\end{itemize}

\textbf{Motivators for Privacy-Protective Practices} (multiple selections allowed):
\begin{itemize}
    \item Technical convenience and ease of use
    \item Being in sensitive locations
    \item Social pressure (avoiding dissatisfaction or conflict)
    \item Respect and moral responsibility
    \item Social incentives (receiving praise or recognition)
    \item Legal and regulatory requirements
    \item Other: \underline{\hspace{3cm}}
\end{itemize}

\section{Study 1 Supplementary Materials}

\subsection{Participant Demographics and Baseline Characteristics}

Table~\ref{tab:demographics} presents the demographic characteristics and baseline attitudes of participants in Study 1.

\begin{table*}[ht]
\centering
\caption{Participant Demographics and Baseline Characteristics}
\label{tab:demographics}
\begin{tabular}{lll}
\toprule
\textbf{Characteristic} & \textbf{Bystanders (N=293)} & \textbf{Wearers (N=232)} \\
\midrule
\textbf{Gender} & & \\
\quad Male & 49.2\% & 59.5\% \\
\quad Female & 50.5\% & 39.2\% \\
\quad Non-binary/Other & 0.3\% & 1.3\% \\
\midrule
\textbf{Age} & & \\
\quad 18--25 years & 39.6\% & 32.3\% \\
\quad 26--35 years & 53.2\% & 51.3\% \\
\quad 36--45 years & 5.1\% & 14.2\% \\
\quad 46+ years & 2.0\% & 2.2\% \\
\midrule
\textbf{Smart Glasses Familiarity} & & \\
\quad Never heard of & 4.8\% & 0.0\% \\
\quad Heard of but unfamiliar & 28.3\% & 13.8\% \\
\quad Understand basic functions & 46.8\% & 35.3\% \\
\quad Very familiar & 15.7\% & 27.6\% \\
\quad Current/former user & 4.4\% & 23.3\% \\
\midrule
\textbf{Acquaintances Using Smart Glasses} & & \\
\quad None & 25.9\% & 28.9\% \\
\quad 1 person & 25.9\% & 24.1\% \\
\quad 2 people & 23.2\% & 18.5\% \\
\quad 3+ people & 24.9\% & 28.4\% \\
\midrule
\textbf{Brand Awareness} & & \\
\quad Xiaomi AI Glasses & 87.7\% & 89.2\% \\
\quad Rayneo V3/X3 Series & 49.1\% & 60.3\% \\
\quad Ray-Ban Meta/Oakley Meta & 27.0\% & 43.5\% \\
\quad Google Glass & 29.7\% & 31.9\% \\
\quad Rokid Glasses & 21.8\% & 34.9\% \\
\midrule
\textbf{Baseline Attitudes} & & \\
\quad Awareness / Perceived Responsibility & 5.76 $\pm$ 1.15 & 5.94 $\pm$ 1.19 \\
\quad Control / Information Sharing Intention & 5.99 $\pm$ 0.93 & 5.75 $\pm$ 1.32 \\
\quad Collection / Privacy Protection Intention & 5.53 $\pm$ 1.19 & 5.98 $\pm$ 1.14 \\
\bottomrule
\end{tabular}
\end{table*}

\subsection{Convergent Validity of Baseline Measures}
\label{appendix:validity}

\begin{figure*}[ht]
\centering
\includegraphics[width=\textwidth]{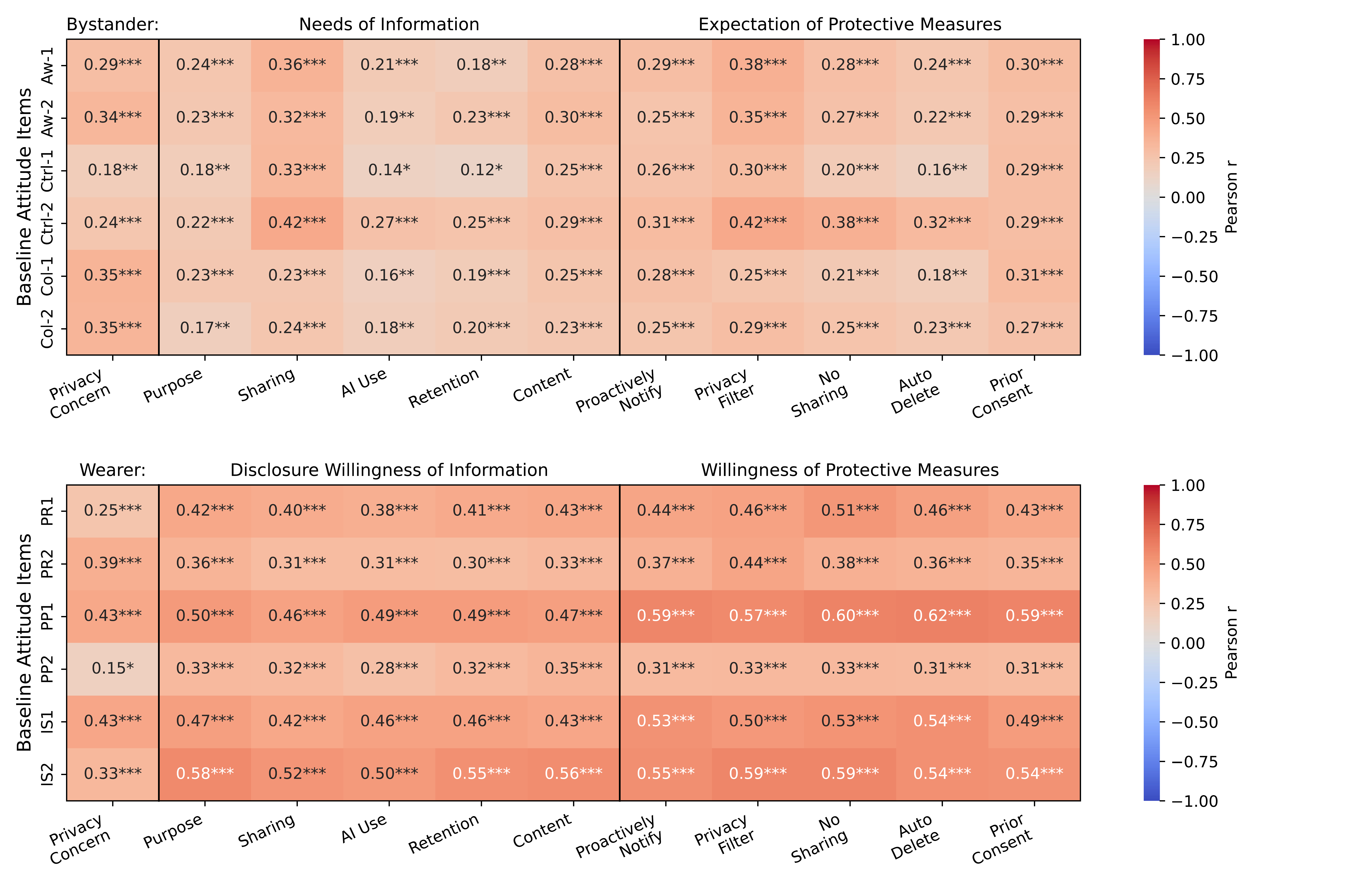}
\caption{Pearson correlations between baseline attitude items and scenario-averaged measures. \textbf{Top panel:} Bystanders' baseline attitudes (Awareness: Aw-1, Aw-2; Control: Ctrl-1, Ctrl-2; Collection: Col-1, Col-2) correlated with their average ratings of privacy concerns, information needs, and expectations for protective mechanisms across scenarios. \textbf{Bottom panel:} Wearers' baseline items (Perceived Responsibility: PR1, PR2; Privacy Protection Intention: PP1, PP2; Information Sharing Intention: IS1, IS2) correlated with their scenario-averaged disclosure willingness and willingness to adopt protective measures. Cells display Pearson's r; asterisks indicate significance levels (* p < .05, ** p < .01, *** p < .001). Darker shading represents stronger positive correlations.}
\label{fig:validity_correlations}
\end{figure*}

To assess whether our abbreviated baseline measures behaved in line with theoretical expectations, we examined their correlations with scenario-based measures (Figure~\ref{fig:validity_correlations}).

For bystanders, all six baseline items showed positive correlations with scenario-averaged privacy concerns, information needs, and protective measure demands (r $\approx$ .12–.42, most ps < .001). Participants reporting higher baseline concern also reported stronger scenario-based privacy concerns and requested more extensive protections.

For wearers, the pattern was more pronounced. Baseline items capturing perceived responsibility, privacy protection intention, and information sharing intention correlated positively with scenario-averaged disclosure willingness and willingness to adopt PETs (r $\approx$ .25–.62, ps < .001). Notably, despite modest internal consistency ($\alpha$ = 0.48), Information Sharing Intention items showed substantial correlations with scenario-based disclosure willingness (r = .42–.59), supporting their validity as descriptive measures.

These patterns confirm convergent validity: baseline measures relate to scenario-specific outcomes in theoretically expected directions. Given their brevity, we treat these scales as descriptive rather than primary constructs.

\end{document}